\documentclass [aps,12pt,a4paper,superscriptaddress]{revtex4}
\usepackage{graphicx}  % needed for figures
\usepackage{subfigure}
\usepackage{float}
\usepackage{dcolumn}   % needed for some tables
\usepackage{bm}        % for math
\usepackage{amsmath}        % for math
\usepackage{amssymb}   % for mathv
\usepackage{natbib}
\usepackage{color}
\begin{document}
%%%%%%%%%%%%%%%%%%%%%%%%%%%%%%%%%%%%%%%%%%%%%%%%%%%%%%%%%%%%%%%%%%%%%
\title{Enhanced scattering induced by electrostatic correlations in concentrated solutions of salt-free dipolar and ionic polymers\footnote{Notice:This manuscript has been authored  by UT Battelle, LLC, under
Contract No. DE6AC0500OR22725 with the U.S. Department of Energy.
The United States Government retains and the publisher, by accepting the
article for publication, acknowledges  that the United States Government
retains a non-exclusive, paid-up, irrevocable, world-wide license to publish
or reproduce the published form of this manuscript, or allow others to do
so, for the United States Government purposes. The Department of
Energy will provide public access to these results of federally sponsored
research in accordance with the DOE Public Access Plan
(http://energy.gov/downloads/doe-public-access-plan).}
}
%%%%%%%%%%%%%%%%%%%%%%%%%%%%%%%%%%%%%%%%%%%%%%%%%%%%%%%%%%%%%%%%%%%%%

\author{Rajeev Kumar}
\email{kumarr@ornl.gov}
\affiliation{Center for Nanophase Materials Sciences, Oak Ridge National Laboratory, Oak Ridge, TN-37831, USA}
\affiliation{Computational Sciences and Engineering Division, Oak Ridge National Laboratory, Oak Ridge, TN-37831, USA}

\author{Bradley Lokitz}
\affiliation{Center for Nanophase Materials Sciences, Oak Ridge National Laboratory, Oak Ridge, TN-37831, USA}

\author{Timothy E. Long}
\affiliation{Department of Chemistry, Virginia Tech., Blacksburg, VA 24061, USA}\affiliation{Macromolecules Innovation Institute (MII), Virginia Tech., Blacksburg, VA 24061, USA}

\author{Bobby G. Sumpter}
\affiliation{Center for Nanophase Materials Sciences, Oak Ridge National Laboratory, Oak Ridge, TN-37831, USA}
\affiliation{Computational Sciences and Engineering Division, Oak Ridge National Laboratory, Oak Ridge, TN-37831, USA}

%%%%%%%%%%%%%%%%%%%%%%%%%%%%%%%%%%%%%%%%%%%%%%%%%%%%%%%%%%%%%%%%%%%%%
\begin{abstract}

\noindent
We present a generalized theory for studying static monomer density-density correlation function (structure factor) in concentrated solutions 
and melts of dipolar as well as ionic 
polymers. The theory captures effects of electrostatic fluctuations on the structure factor and provides insights into the origin of experimentally observed enhanced scattering at ultralow wavevectors in salt-free ionic polymers. It is shown that the enhanced scattering can originate from a coupling between fluctuations of electric polarization and monomer density. Local and non-local effects of the polarization resulting from \textit{finite sized} permanent dipoles and ion-pairs in dipolar and charge regulating ionic polymers, respectively, are considered. Theoretical 
calculations reveal that, similar to the salt-free ionic polymers, the structure factor for dipolar polymers can also exhibit a peak at a finite wavevector and enhanced scattering at ultralow wavevectors. Although consideration of dipolar 
interactions leads to attractive interactions between monomers, the enhanced scattering at ultralow wavevectors is predicted solely on 
the basis of the electrostatics of weakly inhomogeneous dipolar and ionic polymers without considering the effects of any aggregates or phase separation. Thus, we conclude that neither aggregation nor phase separation is necessary for observing the enhanced scattering at ultralow wave-vectors in salt-free 
dipolar and ionic polymers. For charge regulating ionic polymers, it is shown that electrostatic interactions between charged monomers get screened with a screening length, which depends not only on the concentration of ``free'' counterions and coions but also on the concentration of ``adsorbed'' ions on the polymer chains. \textcolor{black}{Qualitative comparisons with the experimental scattering curves for ionic and dipolar polymer melts are presented using the theory developed in this work.} 

\end{abstract}
\date{\today}

\maketitle

\section{Introduction}
Scattering\cite{rayleigh_scattering, einstein_paper,debye_scattering,zimm_molecular,fixman_molecular,zimm_polymer_scattering,kirkwood_lsexpt,correlation_hole,degennes_pincus,higgins_review,daillantbook,forster1995,ise_review,ermi_solvation,ermi_domains,dobrynin_rev,muthu_ls_theory,muthu_ls_expt,muthu_perspective} is one of the most powerful characterization tools for probing structure\cite{fredbook} and dynamics\cite{edwardsbook} of polymers at different length and time scales.
While the protocols for analyzing scattering from neutral polymers are fairly well-established, interpreting scattering from ionic polymers
still poses a great challenge, despite decades of research\cite{kirkwood_lsexpt,forster1995,dobrynin_rev,muthu_ls_theory,muthu_ls_expt,muthu_perspective}. As an example, consider the time-independent (or the static) small-angle
X-ray scattering measurements done on dilute solutions of pH responsive biopolymers, reported as early as 1985 by Matsuoka et al\cite{matsuoka}. Angularly averaged scattering intensity plotted against the magnitude of the wavevector ($q= |\mathbf{q}| = 4\pi\sin \theta/\lambda, \lambda$ and $\theta$ being the wavelength and angle of incidence, respectively, of the wave to be scattered) showed a peak at a finite wavevector and the maximum intensity at lower wavevectors (for $q < 0.02\,$ \AA$^{-1}$). The maximum intensity at the lowest wavevector probed in experiments can differ by two orders of magnitude in comparison with the intensity at the peak corresponding to a local maximum. Despite a number of studies focused on understanding various aspects related to structure and dynamics of polyelectrolytes, the origin of the enhanced scattering at lower wavevectors is still an unsolved puzzle.

The peak in the scattering is now colloquially known as the ``polyelectrolyte peak''
and has become a signature of ionic polymer solutions and melts\cite{muthu_perspective}. The existence of the polyelectrolyte peak in solutions was first 
\textit{conjectured} by de Gennes \textit{et al.}\cite{degennes_pincus} in 1976 and it was argued to originate 
from \textit{purely repulsive interactions}. Their conjecture was based on an analysis of 
angularly averaged monomer density-density correlation function (or the so-called structure factor, defined below) for the polyelectrolyte solutions in the limits of low and high $q$. 
Due to the relevance of their arguments for the problem related to the origin of enhanced scattering at ultralow 
wavevectors, we present those arguments here. 

The scattered intensity by $N_s$ number of scatterers dissolved in a solvent with their local number density written as $c(\mathbf{r})$, is given by\cite{RGD_theory_1,RGD_theory_2}
\begin{eqnarray}
I(\mathbf{q}) &=& I_s(0)\sum_{j=1}^{N_s}\sum_{k=1}^{N_s}\left<c(\mathbf{r}_j)c(\mathbf{r}_k)e^{-i\mathbf{q}\cdot\left[\mathbf{r}_j-\mathbf{r}_k\right]}\right> = I_s(0)V^2\left<\tilde{c}(\mathbf{q})\tilde{c}(-\mathbf{q})\right>  = I_s(0) N_s S(\mathbf{q})\label{eq:rayleigh} 
\end{eqnarray}
where the prefactor $I_s(0)$ carries the information about the scattering geometry, scattering volume and the nature of interactions between the incident wave and material. Here, $\mathbf{q}$ is the scattering wavevector, $\tilde{c}(\mathbf{q})$ is the Fourier component of $c(\mathbf{r})$, $V$ is the volume and Eq. ~\ref{eq:rayleigh} acts as a definition of the
structure factor \textit{per scatterer}, $S(\mathbf{q})$. In the limit of low $q = |\mathbf{q}|$ and in particular, for $q=0$, $S(0) = k_B T (\partial^2 F_{sol}/\partial c^2)^{-1}$ (a relation 
first derived by Einstein\cite{einstein_paper}), where $F_{sol}$ is the free energy of the solution containing the scatterers as solutes, $k_B$ is the Boltzmann 
constant and $T$ is the temperature. The partial derivative needs to be computed at uniform concentration of the scatterers, i.e., at $c \equiv c_h = N_s/V$, where the subscript $h$ means homogeneous. The relation between the structure factor and the changes in the 
free energy is formally exact for solutions exhibiting small fluctuations in
the monomer number densities\cite{einstein_paper}. de Gennes \textit{et al.}\cite{degennes_pincus} wrote this relation 
in terms of the osmotic pressure contribution due to the polyelectrolyte chains ($\Pi_p$ in their notation) so that 
$S(0) = k_B T (\partial \Pi_{p}/\partial c)^{-1}$, where $c$ is the number density of the chains. 
Furthermore, de Gennes \textit{et al.} expected $\Pi_p \sim c k_B T$ so that $S(0)$ is a constant, which was  
\textit{assumed} to be  of order unity without constructing any microscopic model to compute the osmotic pressure. For high $q$, so that correlations inside a segment are being probed, it was argued that scattering from \textit{a single chain} should be $S_1(q) =  \pi/qb, b$ being the length of a segment along the chain. The inverse dependence on $q$ highlights a fractal dimension of unity and rodlike nature of the charged chain as well as the segments. Inspired by the success of blob arguments for the neutral polymers and noticing that different 
segments don't overlap with each other, 
it was conjectured that scattering from many chains should be the same as a single chain i.e., $S(q) =  S_1(q) = \pi/qb = \pi g/q \xi$, where $g$ is the number of monomers/segments in each blob and
$\xi = g b$ is the length of a blob. $\xi$ was called the concentration dependent
correlation length by de Gennes \textit{et al}. If one constructs an inperpolation function, which satisifes the limits of low and high 
$q$ then the function will go through a maximum at $q\sim 1/\xi$. The maximum appears due to the assumption of $g\gg 1$, which also makes the entropic contribution from the chains to the osmotic pressure minuscule. Such an analysis led to the suggestion that there must be a peak 
in the structure factor as well as the scattering intensity. Since then, dependencies of the peak on the concentration of polymers, salt concentration,
temperature etc., have been studied extensively and are very well documented in the literature\cite{ise_review,ermi_solvation,ermi_domains,Muthukumar2016}. It has been established that the structure factor of the polyelectrolytes \textcolor{black}{does not} necessarily follow $1/q$ behavior at high $q$ and in fact, its scaling with $q$ depends on the concentration of chains and salt ions. The most successful theory to-date in explaining dependencies of the polyelectrolyte peak on various experimental variables is the double screening theory, developed by Muthukumar\cite{muthu_double_screening,muthu_perspective}, which is based on the concepts of screening by small ions (counterions and coions) and chains. The double screening theory describes the origin of the polyelectrolyte peak solely on the basis of correlations among monomers in a homogeneous medium i.e., without considering effects of any aggregates which may or may not be present. Furthermore, in contrast to the de Gennes \textit{et al.} arguments, density fluctuations in \textit{finite sized} polymers\cite{muthu_double_screening,muthu_perspective} were shown to cause attraction between similarly charged monomers. 
However, the double screening theory \textcolor{black}{does not} predict enhanced scattering at ultralow wavevectors in solutions of salt-free ionic polymers\cite{Muthukumar2016}. In this work, we will show that this limitation of the theory results from treatment of solvent as a uniform dielectric continuum and neglecting 
effects of charge regulation as well as fluctuations in the electric polarization. 

The enhanced scattering intensity at ultralow wavevectors has been interpreted by invoking the idea of aggregation\cite{ermi_domains} in polyelectrolyte solutions. In particular, it is \textit{assumed} that there is an \textit{attraction} between similarly charged polymers leading to long-lived
aggregates, whose scattering dominates over the scattering of individual chains at low wavevectors. The concept of aggregation in
solutions of charged polymers is supported by the time-dependent (or the dynamic) scattering measurements\cite{ermi_domains,sedlak_dynamics}, which show at least two diffusive
modes, called the fast and the slow modes. Physically, multiple diffusive modes highlight the dynamic heterogenity in ionic polymer solutions. Typically, the fast and the slow modes are interpretted as the diffusion
of single chains and the aggregates, respectively. In seminal works spanning almost two decades, Muthukumar\cite{muthu_double_screening,muthu_pnas,muthu_perspective} has shown that
like-charge attraction can originate from the density fluctuations and described the concentration dependencies of the diffusion coefficients for the fast and the slow modes by taking into account the
effects of dipoles originating from adsorbed counterions on the polymers. 

In addition to the highly non-trivial and counterintuitive 
notion of like-charge attraction, the possibility of describing an enhanced scattering without invoking the idea of aggregation 
is worth exploring. Indeed, aggregation can lead to excess scattering\cite{Muthukumar2016,muthu_perspective} but it is not clear if the aggregation is a necessary condition for observing an excess scattering at ultralow wavevectors. For example, aggregation can lead to a peak at 
a finite wavevector\cite{yarusso_model,ionomer_upturn,winey_review} in the angularly averaged scattered intensity, where the peak position 
characterizes the avergage length scale of the aggregates. However, the peak \textcolor{black}{need not} be described by invoking aggregation always and can be described solely on the basis of correlations among connected monomers, 
as shown by de Gennes\cite{correlation_hole} in a treatement of the so-called 
``correlation hole'' using the random phase approximation (RPA) for neutral chains and by a number of researchers\cite{borue_1988,joanny_1990,vilgis_rpa,liverpool_counterions,Muthukumar2016,muthu_perspective} for the polyelectrolyte chains. Furthermore, qualitatively similar features including
a peak at finite wavevector and enhanced scattering intensity at lower wavevectors have been
observed in experiments probing structure of ionomers\cite{yarusso_model,ionomer_upturn,winey_review} (i.e., weakly charged polymer melts), dipolar polymers like poly(ethyleneoxide) in deuterated water\cite{hammouda_peo_sans} using small angle neutron scattering and zwitterionomers\cite{long_paper} (i.e., melts of zwitterionic polymers) using X-rays. 
As the enhanced scattering is typically observed in salt-free ionic and dipolar polymers, where
electrostatic effects are significant, it is expected that electrostatic interactions are responsible for the enhanced scattering.

In this work, we focus on finding an origin of the enhanced scattering at ultralow wavevectors on the basis of electrostatic
interactions without considering any kind of aggregation. For such purposes, we have constructed a minimal model 
for the monomer density-density correlation function (static structure factor), which shows conclusively that \textit{aggregation is not a necessary condition for observing the enhanced scattering and the peak in the cases of salt-free ionic and dipolar polymers}. \textcolor{black}{The model is minimal in the sense that effects of charge regulation and polarization are considered without considering additional effects due to semi-flexible backbones, finite polarizability of monomers and complications arising from hydrogen bonding in polymers as well as solvents such as water.} Analysis of the monomer density-density correlation function reveals that dipolar interactions can lead to 
 the enhanced scattering. Dipolar interactions tend to lower the osmotic pressure due to attraction and lead to additional wave-vector dependence in the monomer density-density 
correlation function. For the case of ionic polymers with added salt or solvent molecules, we
integrate out the degrees of freedom of the counterions, co-ions and solvent to obtain an \textit{effective} scattering function per monomer/segment. Analytical expressions for the static structure factor in concentrated solutions and melts containing dipolar and salt-free ionic polymers are derived. These expressions are based on RPA\cite{borue_1988,joanny_1990,vilgis_rpa,liverpool_counterions} for understanding the effects of dielectric inhomogeneity,
charge regulation/ion-pairing and ion-dipole interactions. We consider only the high-temperature regime of rotating dipoles (i.e., weak-coupling limit for the dipoles) without considering
any frustrated states, similar to our previous works related to dipolar effects in polymeric systems\cite{mahalik_2017,mahalik_2016,kumar_muthu}. Polymer segments and solvent molecules are assumed to have fixed permanent dipoles embedded in a finite volume characterized by a length
scale\cite{mahalik_2017,mahalik_2016}. Charge regulation due to the counterion adsorption is considered using a two state model similar to our previous work on the pH responsive polyelectrolytes\cite{kumar_kilbey}. Predictions for monomer density profiles based on the two state model have been compared with the experimental density profiles determined using neutron
reflectivity profiles and excellent agreements were found\cite{mahalik_pe_brushes_2016}. In this paper, we show that considerations of the charge regulation within the two state model along with the electrostatics of finite sized
ions and ion-pairs/dipoles can lead to a structure factor exhibiting a peak at a finite wavevector and enhanced scattering at ultralow wavevectors. \textcolor{black}{Finite size of dipoles as well as ions and dipolar interactions are shown to be the primary cause of such a shape of the structure factor-wavevector curve. However, charge regulation and its coupling with 
polarization fluctuations are considered to present a more realistic description of charged polymers.} We should again emphasize that the 
enhanced scattering described in this work originates without invoking any phase segregation or aggregation and arises solely from the electrostatics of finite sized ions and ion-pairs, where the latter is shown to be equivalent to a non-local dielectric medium. Furthermore, it will be shown that the electrostatic fluctuations for charge regulating polymers along with dipolar interactions lead to attractive interactions between charged monomers similar to classic works by Kirkwood and Shumaker\cite{kirkwood_fluctuations, kirkwood_forces}. Before presenting the mathematical details, we derive some of the results in a heuristic manner, with an intent that it will provide a clearer perspective on the origin of the enhanced scattering.

This paper is organized as follows. In the next section, the main results for salt-free melts are discussed by using a heuristic approach, which is developed on the basis of a detailed mathematical analysis. The mathematical analysis is presented in the section ~\ref{sec:construct}, which leads to an effective one-component description for the salty charge regulating polymer solutions. Free energies and monomer density-density correlation functions are presented in the section ~\ref{sec:results}. Comparisons with small angle X-ray scattering experiments on salt-free dipolar and ionic polymer melts are also presented in the section ~\ref{sec:results}. Conclusions and future directions are presented in the section ~\ref{sec:conclusions}.

\section{Heuristic Approach}\label{sec:heuristic}
First, we consider melts containing dipolar homopolymers. For weakly inhomogeneous melts, 
fluctuations in the local volume fraction of monomers ($c_p(\mathbf{r})\equiv c(\mathbf{r})/c_o$ so that 
$c_o$ is the spatially averaged number density of monomers) and angularly averaged electric polarization ($P(\mathbf{r})$) contribute to the probability distribution for realizing configurations with prescribed inhomogeneities. These contributions can be written in the form of an effective Hamiltonian (see the next section for the derivation) as
\begin{eqnarray}
\frac{H_{eff}}{k_B T} &=& \frac{c_o^2}{2}\int d\mathbf{r} \int d\mathbf{r}'c_p(\mathbf{r})g_p^{-1}(|\mathbf{r}-\mathbf{r}'|)c_p(\mathbf{r}') + \frac{w_{pp}c_o^2}{2}\int d\mathbf{r} \rho_p^2(\mathbf{r}) \nonumber \\
&& + \frac{1}{2\lambda_0}\int d\mathbf{r} P^2(\mathbf{r}) + \frac{1}{2\lambda_1}\int d\mathbf{r} \left[\nabla_{\mathbf{r}}P(\mathbf{r})\right]^2 \label{eq:heuristic_heff}
\end{eqnarray}
The first term captures effects of the chain connectivity and $g_p^{-1}(|\mathbf{r}-\mathbf{r}'|)$ is the pair 
correlation function for the chains in the absence of any interactions. 
The second term arises from the excluded volume interactions, whose strength is characterized by the parameter $w_{pp}$. However, instead of having a standard 
Edwards's\cite{edwardsbook} point-like interaction range, these interactions are introduced by smearing
the monomer density about the center of mass of the monomers so that a length scale, $a_p$, appears in the description, which characterizes the range of the smearing. In particular, we consider
\begin{eqnarray}
\rho_p(\mathbf{r}) &=& \int d\mathbf{r}'\hat{h}_p(\mathbf{r}-\mathbf{r}')c_p(\mathbf{r}')
\end{eqnarray}
where $\hat{h}_p(\mathbf{r}) = \exp(-\pi |\mathbf{r}|^2/2a_p^2)/(2a_p^2)^{3/2}$ is one of the physically motivated and mathematically convenient choices. Physically, $a_p$ is the length scale, which captures the effects of finite 
size of the monomers and can also represent the size of short side groups on a monomer. The third and the fourth 
terms in Eq. ~\ref{eq:heuristic_heff} take into account local and non-local effects of the polarization, respectively. For the dipolar polymers, the polarization and the monomer density are coupled. The simplest 
relation for the coupling between the polarization and the density can be derived by generalizing the Langevin-Debye model, originally derived for a 
homogeneous medium\cite{dielectric}, to an inhomogeneous medium\cite{kumar_kilbey}. Such a generalization leads to a linear relation $P(\mathbf{r}) = \Delta_p \rho_p(\mathbf{r})$, where 
$\Delta_p = 4\pi l_{Bo}p_p^2 c_o/3$ so that 
$l_{Bo} = e^2/4\pi \epsilon_o k_B T$ is the Bjerrum length in vacuum, $e$ is the charge of an electron, $\epsilon_o$ is 
the permittivity of vacuum and $p_p$ is the length of the dipole on the monomer. $\lambda_0$ and $\lambda_1$ are the molecular parameters, which control the magnitude and range of 
local and non-local effects of the polarization, respectively. As the dipolar interactions are 
attractive and the local effects of the polarization\cite{kumar_glenn} can be merged with the excluded volume 
interaction term in Eq. ~\ref{eq:heuristic_heff}, in general, $\lambda_0 \sim - a_p^3 < 0$. 
In contrast, spatial gradients of the polarization (non-local effects) tend to cost energy\cite{kumar_muthu,mahalik_2016} and $\lambda_1 \sim a_p > 0$.

Using the Fourier transform and representing transformed variables with superscript \textasciitilde, Eq. ~\ref{eq:heuristic_heff} can be rewritten as
\begin{eqnarray}
\frac{H_{eff}}{k_B T} &=& \frac{V}{2}\sum_{\mathbf{q}}\left[c_o^2 \tilde{g}_p^{-1}(q) 
+ \left\{w_{pp} c_o^2 + \Delta_p^2\left(\frac{1}{\lambda_0} + \frac{q^2}{\lambda_1}\right)\right\}\tilde{h}_p(q)\tilde{h}_p(-q)\right]\tilde{c}_p(\mathbf{q})\tilde{c}_p(-\mathbf{q}) \label{eq:heuristic_heff_FT}
\end{eqnarray}
and leads to  
\begin{eqnarray}
\left\langle\tilde{c}_p(\mathbf{q})\tilde{c}_p(-\mathbf{q})\right\rangle_{H_{eff}} &=& \frac{1}{V\left[c_o^2 \tilde{g}_p^{-1}(q) 
+ \left\{w_{pp} c_o^2 + \Delta_p^2\left(\frac{1}{\lambda_0} + \frac{q^2}{\lambda_1}\right)\right\}\tilde{h}_p(q)\tilde{h}_p(-q)\right]} \label{eq:heuristic_dip_st}
\end{eqnarray}
where the average in Eq. ~\ref{eq:heuristic_dip_st} is evaluated using the Boltzmann distribution weighted by $\exp(-H_{eff}/k_B T)$. For $n_p$ monodisperse 
flexible chains, each containing $N$ Kuhn segments of equal lengths $b$, in a volume $V$, $\tilde{g}_p^{-1}(q) = 1/(c_o Ng_D(q^2 Nb^2/6))$, where $c_o = n_pN/V$ and 
$g_D(x) = 2(e^{-x}-1+x)/x^2$ is the Debye function\cite{debye_scattering}. Also, $\tilde{h}_p(q) = \exp(-q^2 a_p^2/2\pi)$ for $\hat{h}_p(\mathbf{r}) = \exp(-\pi |\mathbf{r}|^2/2a_p^2)/(2a_p^2)^{3/2}$. Using Eqs. ~\ref{eq:rayleigh} and ~\ref{eq:heuristic_dip_st}, with $N_s = n_pN$, we can write   
\begin{eqnarray}
I(\mathbf{q}) &=& I_s(0) \left[c_o^2 V^2\right]\left<\tilde{c}_p(\mathbf{q})\tilde{c}_p(-\mathbf{q})\right>_{H_{eff}} \nonumber \\
&=& \frac{I_s(0)c_o V}{\frac{1}{Ng_D(q^2 Nb^2/6)} + \left\{w_{pp} c_o + \frac{\Delta_p^2}{c_o}\left(\frac{1}{\lambda_0} + \frac{q^2}{\lambda_1}\right)\right\}\tilde{h}_p(q)\tilde{h}_p(-q)} \equiv I(q) \label{eq:heuristic_I}
\end{eqnarray}
Here, $I(\mathbf{q}) \equiv I(q)$ i.e., the scattering intensity only depends on
magnitude of the wavecector due to the use of angularly averaged interaction potential and polarization in Eq. ~\ref{eq:heuristic_heff}. Such a functional form for the scattering intensity provides three important insights. First, local 
effects of the polarization are found to renormalize the bare excluded 
volume parameter so that 
\begin{eqnarray}
w_{pp,r} &=& w_{pp} + \frac{\Delta_p^2}{c_o^2 \lambda_0} \label{eq:renorm_wpp}
\end{eqnarray} 
As $\lambda_0 <0$, 
the dipolar interactions tend to decrease the renormalized excluded volume parameter. This is in agreement with 
treatment of freely rotating dipoles\cite{kumar_glenn}. Intensity at 
$q=0$ is given by (cf. Eq. ~\ref{eq:heuristic_I})
\begin{eqnarray}
I(0) &=& \frac{I_s(0)c_o V}{\frac{1}{N} + w_{pp,r} c_o } \label{eq:heu_I0}
\end{eqnarray}
and depends on the parameters $I_s(0)c_o N V$ and $w_{pp,r} c_o N$.
As the dipolar interactions tend to decrease $w_{pp,r}$ (cf. Eq. ~\ref{eq:renorm_wpp}), 
$I(0)$ is predicted to increase with an increase in magnitude of $\frac{\Delta_p^2}{c_o^2 \lambda_0} \sim \frac{l_{Bo}p_p^2}{c_o a_p^3}$. \textcolor{black}{It should be noted that $w_{pp,r} c_o = 0$ corresponds to the stability limit of the homogeneous phase\cite{zimm_polymer_scattering} and macrophase separation can occur when $w_{pp,r} c_o < 0$. Eq. ~\ref{eq:heu_I0} shows that the macrophase separation can appear in the form of divergent scattering at the zero wave-vector for infinitely long polymers i.e., $I(0)\rightarrow \infty$ becomes a signature of the macrophase separation. In this work, we consider homogeneous phases so that $w_{pp,r} c_o \geq 0$.}

\begin{figure}[ht]
  \includegraphics[width=0.65\linewidth]{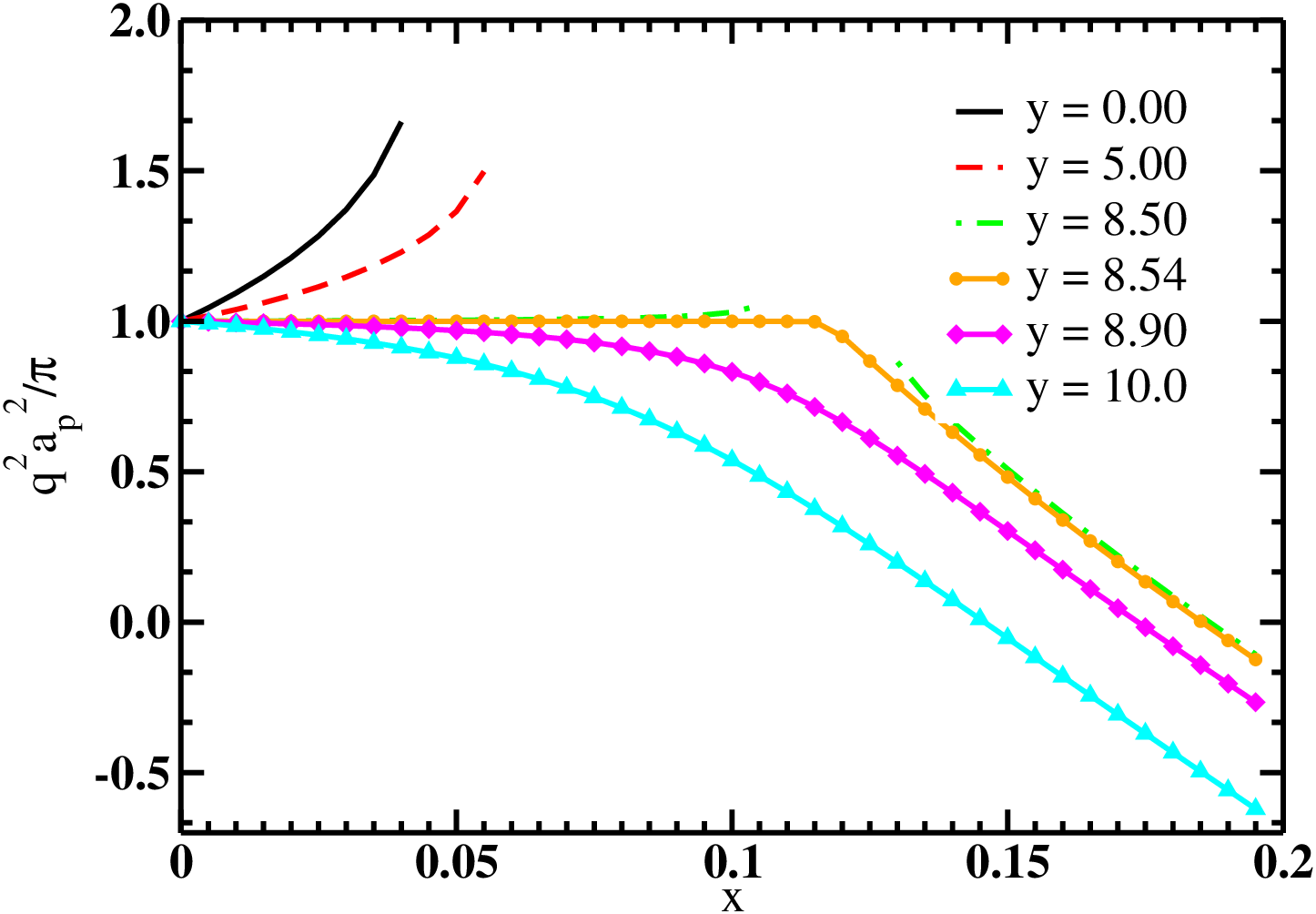}
\caption{Effects of non-local effects of the electric polarization (characterized by $x = \beta_0 b^2/6\zeta_d^2$) on location of the local
minimum in the scattering intensity for dipolar polymer melts. Plots presented here are obtained from Eq. ~\ref{eq:nonzero_q}. The right hand side of Eq. ~\ref{eq:nonzero_q} becomes a complex number for higher values of $x$ when $y<8.5$. In contrast, the right hand side of Eq. ~\ref{eq:nonzero_q} 
becomes negative for higher values of $x$ when $y>8.5$. \label{fig:heu_dip_effects}}
\vspace{0.2in}
\end{figure}

\begin{figure}[ht]  \centering
\subfigure[]{\includegraphics[width=0.65\linewidth]{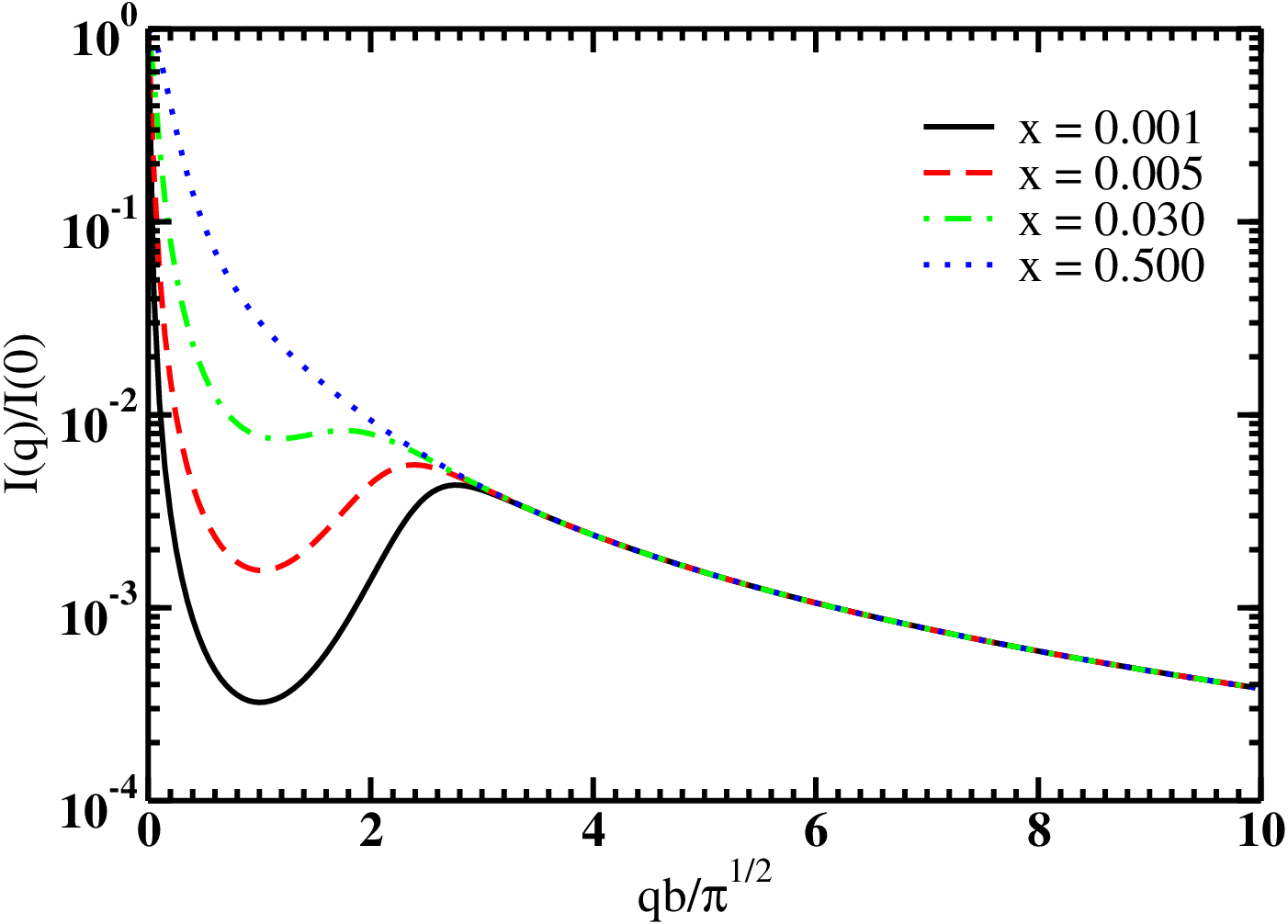}}
\hspace{0.1in}\subfigure[]{\includegraphics[width=0.65\linewidth]{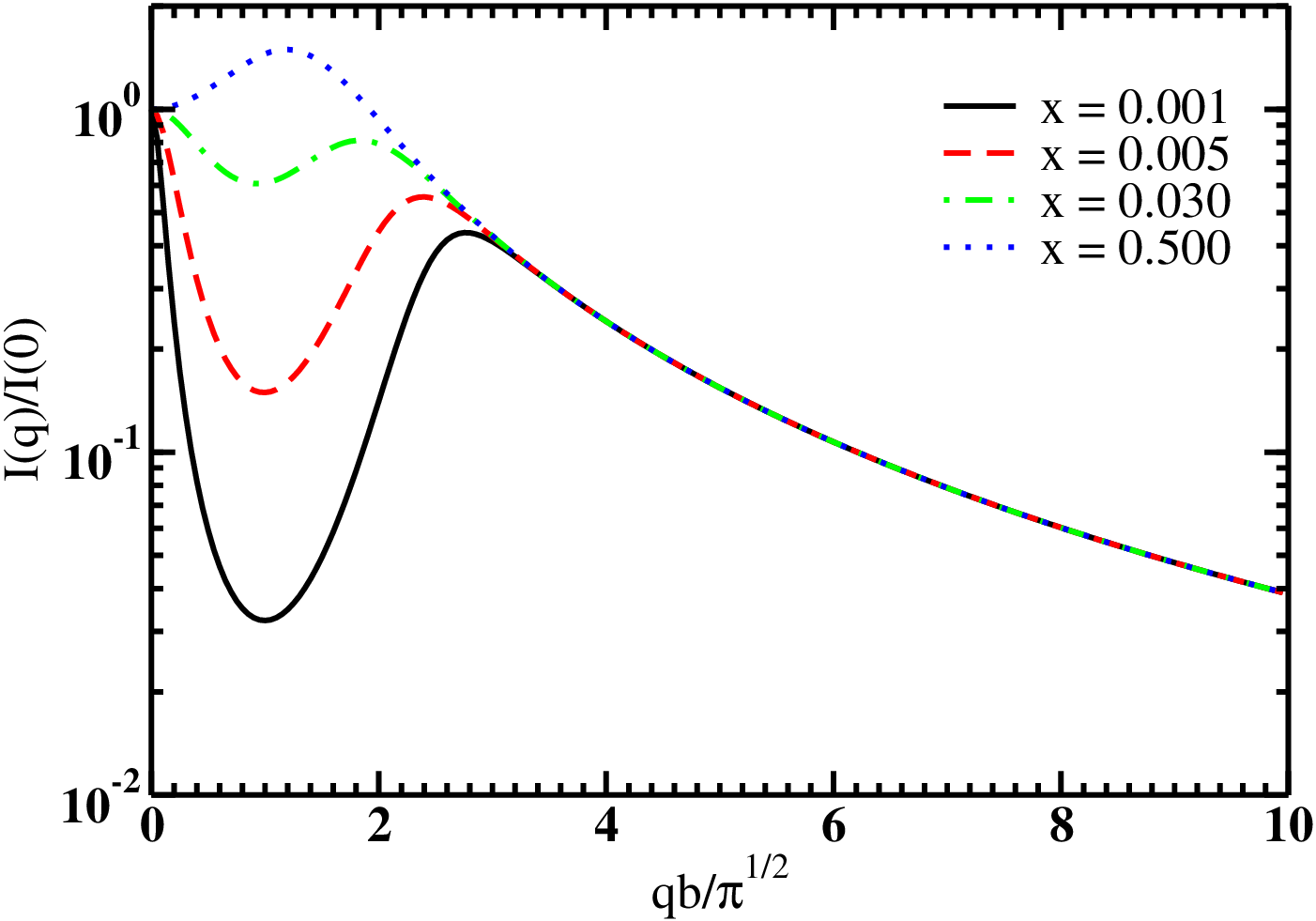}}
\caption{Non-local effects of the polarization, characterized by $x = b^2/12 \zeta_d^2$, on the scattering intensity from melts containing dipolar polymers. (a) $w_{pp,r}c_o = 0, a_p/b = 1, N = 100 \,(i.e., y=0)$. (b) $w_{pp,r}c_o = 1, a_p/b = 1, N = 100 \, (i.e., y=12)$.\label{fig:intensity_dipolar_effects}}
\vspace{0.2in}
\end{figure}

Second, $I(q)$ is a \textit{non-monotonic} function of $q$ and the origin of the non-monotonicity 
can be readily identified if we optimize the function, $I_s(0)/I(q)$, with 
respect to $q$. Such an optimization can be done analytically if we use 
an approximation $g_D(x) \simeq 1/(1 + \beta_0 x), \beta_0 \simeq 0.5$. The approximation for the Debye function leads to a maximum of $\sim 15 \%$ error\cite{edwardsbook} at intermediate values of $x$. Analytical calculations reveal that there is a global maximum in $I(q)$ 
at $q=0$ and a local minimum appears for a non-zero value of $a_p$ at 
\begin{eqnarray}
\frac{q^2 a_p^2}{\pi} &=& (1-xy)-W\left[-\pi x \exp\left(1-xy\right)\right] \label{eq:nonzero_q}\\
x &=& \frac{\beta_0 b^2}{6\zeta_{d}^2},\qquad \zeta_d^2 = \frac{\pi \Delta_p^2}{\lambda_1 c_o}\label{eq:zetad}\\
y &=& \frac{a_p^2}{\zeta_{ed}^2}, \qquad \zeta_{ed}^2 = \frac{\beta_0 b^2}{6 w_{pp,r} c_o}\label{eq:zetaed}
\end{eqnarray} 
Here, $W(x) = \sum_{n=1}^{\infty}(-1)^{n-1}x^n n^{n-2}/(n-1)!$ is the Lambert W-function\cite{Corless1996}. In Eqs. ~\ref{eq:zetad} and ~\ref{eq:zetaed}, we have defined two length scales $\zeta_d$ and $\zeta_{ed}$, which characterize the scale of inhomogeneities in the polarization and concentration, respectively. In particular, $\zeta_{ed}$ is the Edwards's correlation length\cite{edwardsbook,fredbook}. Eq. ~\ref{eq:nonzero_q} reveals that a condition for the existence of the local minimum 
in the scattering intensity is $(1-xy)-W\left[-\pi x \exp\left(1-xy\right)\right] > 0$. Furthermore, Eq. ~\ref{eq:nonzero_q} shows that the local minimum appears at $q = \sqrt{\pi}/a_p$ for $x\rightarrow 0$. Eq. ~\ref{eq:nonzero_q} is plotted in Fig. ~\ref{fig:heu_dip_effects} for 
different values of $x$ and $y$. It is found that location of the 
local minimum can shift to either lower or the higher value of $q$ with an increase in $x$. For small value of $y$, the local minimum shifts to higher values before giving way to a 
monotonic scattering curve. In contrast, for higher values of $y\geq 8.54$, the location of the minimum stays invariant to variations in $x$ before shifting to lower values and eventually appearing 
as a global maximum in the scattering intensity at a non-zero $q$. Analytical calculations show that $\pi x \exp\left[(1-xy)-W\left[-\pi x \exp\left(1-xy\right)\right]\right] > 1$ is required to have a global maximum at a non-zero $q$. Numerical calculations based on Eqs. ~\ref{eq:heuristic_I} and ~\ref{eq:heu_I0} confirm such an interplay of concentration fluctuations (characterized by $y$) and non-local effects of the polarization (characterized by $x$) as shown in Fig. ~\ref{fig:intensity_dipolar_effects}. In Fig. ~\ref{fig:intensity_dipolar_effects}(a) obtained for $y = 0, a_p/b = 1$, it is shown that the non-local effects of the polarization can lead to a non-monotonic scattering intensity as a function of $q$ with a local minimum appearing at $qb/\sqrt{\pi} = 1$, which is in agreement with Fig. ~\ref{fig:heu_dip_effects}. For non-zero values of 
$y$, interplay of the concentration fluctuations and non-local effects of the polarization can lead to a shift of the global maximum from $q=0$ to a non-zero value of $q$ (cf. Fig. ~\ref{fig:intensity_dipolar_effects}(b)). The appearance of a peak at a non-zero $q$ in the structure factor 
can be interpreted using the ``correlation hole'' picture developed by de Gennes\cite{correlation_hole} for incompressible polymer melts. According to the correlation hole picture, the radial distribution function is minimum at the center of mass of a monomer and peaks at a finite distance from the monomer due to repulsive 
interactions. For incompressible polymer melts, the radial distribution function at the center of mass of the monomer is much lower than its value far from the monomer, which leads to almost zero scattering intensity at $q=0$. In contrast, here we have considered a compressible polymer melt which leads to a finite non-zero scattering intensity at $q = 0$. Furthermore, we should point out that larger values of $y$ requires larger values of $a_p/b$ in addition to increased values of excluded volume interactions, $w_{pp,r}c_o$. 
Larger values of $a_p$ can be realized in
experiments by placing dipolar groups on the side chains such as in the case of 
zwitterionomers\cite{long_paper}. 

The third insight obtained from Eq. ~\ref{eq:heuristic_I} is related to the pair 
correlation function, defined by 
\begin{eqnarray}
\left<c(0)c(\mathbf{r})\right> - c_o^2 &=& g_2(r) = \frac{c_o}{2\pi^2}\int_0^{\infty} dq q^2 S(q) \frac{\sin qr}{qr} \label{eq:pair_corr}\\
S(q) &\simeq& \frac{N(1 + q^2 a_p^2/\pi)}{(1 + q^2 a_p^2/\pi)(1+ \beta_0 q^2 R_{g0}^2) + w_{pp,r}c_o N} \label{eq:pair_corr_2}
\end{eqnarray}
where $R_{g0}^2 = Nb^2/6$ and we have used Eq. ~\ref{eq:rayleigh}. In deriving Eq. ~\ref{eq:pair_corr_2}, we have neglected the non-local effects of the polarization to evaluate the Fourier transform analytically and used an additional approximation $\tilde{h}_p(q) \simeq 1 /(1 + q^2 a_p^2/2\pi)$. The approximation for $\tilde{h}_p(q)$ restricts the analytical calculations to small values of $a_p$. Evaluating the integral in Eq. ~\ref{eq:pair_corr}, we get
\begin{eqnarray}
g_2(r) &=& \frac{c_o N}{4\pi (\sqrt{\beta_0}R_{g0})^3}\left[\left(\frac{1}{d^2 \hat{a}_p^2}-\cos 2 t \right)\frac{\sin [(d \sin t)\bar{r}]}{\sin 2t} + \cos [(d \sin t) \bar{r}]\right]\frac{\exp\left[-(d\cos t)\bar{r}\right]}{\bar{r}} \nonumber \\
&& \label{eq:g2r}
\end{eqnarray}
where
\begin{eqnarray} 
d &=& \frac{\left(1 + w_{pp,r}c_o N\right)^{1/4}}{\sqrt{\hat{a}_p}}\\
d\cos t &=& \sqrt{\frac{d^2}{2} + \frac{1 + \hat{a}_p^2}{4\hat{a}_p^2}}\\ 
d\sin t &=& \sqrt{\frac{d^2}{2} - \frac{1 + \hat{a}_p^2}{4\hat{a}_p^2}}
\end{eqnarray} 
Here, we have defined $\hat{a}_p = a_p/\sqrt{\pi \beta_0} R_{g0}$ and $\bar{r} = r/\sqrt{\beta_0} R_{g0}$. The oscillatory nature of the pair correlation function can be immediately seen from Eq. ~\ref{eq:g2r}, which exists for non-zero values of $a_p$ only. \textcolor{black}{For $a_p = 0$,  Eq. ~\ref{eq:pair_corr} gives $g_2(r) =  c_o N \exp\left[-\left(1 + w_{pp,r}c_o N\right)^{1/2}\bar{r}\right]/(4\pi (\sqrt{\beta_0}R_{g0})^3\bar{r})$, which is monotonically decaying function of $r = |\mathbf{r}|$ and identical to the pair correlation function derived by Edwards\cite{edwardsbook} in the limit of $N\rightarrow \infty$.} In other 
words, the finite size of the monomers leads to oscillatory pair distribution function, which is in agreement with works by Kirkwood\cite{kirkwood_stat,kirkwood_review}, Kjellander\cite{dressed_ions} and Muthukumar\cite{muthu_double_screening}.

For salt-free charge regulating polymers containing counterions (denoted by subscript $c$), an effective Hamiltonian similar to Eq. ~\ref{eq:heuristic_heff} can be written by considering charge-charge interactions along with the excluded volume interactions and the interactions among ion-pairs resulting from adsorption of the counterions. Considering a two-state model for the charge regulation and the Langevin-Debye model, 
$P(\mathbf{r}) = \Delta_p (1-\alpha_p) \rho_p(\mathbf{r})$, where
$\alpha_p$ is the degree of ionization of the segments so that $(1-\alpha_p) \rho_p(\mathbf{r})$ is the local volume fraction of ion-pairs. For weakly inhomogeneous melts of charge regulating polymers, charge-charge interactions get screened by the presence of other ions and facilitate formulation of an effective Hamiltonian. However, smearing charges over a finite volume similar to the dipoles leads to wave-vector dependent dielectric and screening effects (see the next section for the details). 
In particular, an effective Hamiltonian for charge regulating polymers can 
be written as 
\begin{eqnarray}
\frac{H_{eff}}{k_B T} &=& \frac{c_o^2}{2}\int d\mathbf{r} \int d\mathbf{r}'c_p(\mathbf{r})g_p^{-1}(|\mathbf{r}-\mathbf{r}'|)c_p(\mathbf{r}') + \frac{w_{pp}c_o^2}{2}\int d\mathbf{r} \rho_p^2(\mathbf{r}) \nonumber \\
&& + \frac{1}{2\lambda_0}\int d\mathbf{r} P^2(\mathbf{r}) + \frac{1}{2\lambda_1}\int d\mathbf{r} \left[\nabla_{\mathbf{r}}P(\mathbf{r})\right]^2 \nonumber \\
&&  + \frac{4\pi l_{Bo}z_p^2 \alpha_p^2 c_o^2}{2}\int d\mathbf{r} \int d\mathbf{r}'\rho_p(\mathbf{r})U_{pp}^{-1}(|\mathbf{r}-\mathbf{r}'|)\rho_p(\mathbf{r}')\label{eq:heuristic_heff_charge}
\end{eqnarray} 
where $z_p$ is the valency of the monomers. $U_{pp}^{-1}(|\mathbf{r}-\mathbf{r}'|)$ is the electrostatic pair-interaction potential for interactions in non-local dielectric medium and screened environment. In particular, it will be shown that $U_{pp}^{-1}(|\mathbf{r}-\mathbf{r}'| = \sum_{\mathbf{q}}\tilde{U}_{pp}(q)e^{i\mathbf{q}\cdot (\mathbf{r}-\mathbf{r}')}$, where 
\begin{eqnarray}
\tilde{U}_{pp}(q) &=& \frac{1}{q^2 \tilde{\epsilon}(q) + \tilde{\kappa}_o^2(q)} - 2\pi l_{Bo}z_p^2(1-\alpha_p)^2\tilde{M}(q)\label{eq:vq}\\
\tilde{\kappa}_o^2(q) &=& 4\pi l_{Bo}z_p^2 \alpha_p \left[2-\alpha_p \right]c_o\tilde{h}_p(q)\tilde{h}_p(-q) \label{eq:screenq}\\
\tilde{\epsilon}(q) &=& 1 + \frac{4\pi l_{Bo}p_p^2 (1-\alpha_p)c_o}{3}\tilde{h}_p(q)\tilde{h}_p(-q) \label{eq:epsilonq}\\
\tilde{M}(q) &=& \frac{1}{16 q \left[\tilde{\epsilon}(0)\right]^2} \left[1-\frac{2}{\pi}\arccos\left(\frac{q}{\sqrt{q^2 + \tilde{\kappa}_o^2(0))}}\right) \right . \nonumber \\
&& \left . + \frac{2}{\pi} \arctan\left(\frac{q^3}{\tilde{\kappa}_o(0)\left\{3q^2 + 4\tilde{\kappa}_o^2(0)\right\}} \right) \right] \label{eq:capM_saltfree}
\end{eqnarray}
In general, $\lambda_0$ and $\lambda_1$ increase with an increase in $\tilde{\kappa}_o^2(q)$.
Implications of the additional charge-charge interactions on the scattering intensity can be seen if one follows the same steps as presented above for dipolar polymers
in getting Eq. ~\ref{eq:heuristic_I}, so that  
\begin{eqnarray}
I(q) &=& \frac{I_s(0)c_o V}{\frac{1}{Ng_D(q^2 Nb^2/6)} + \left\{w_{pp,r} c_o + \frac{\Delta_p^2 (1-\alpha_p)^2 q^2}{c_o \lambda_1} + 4\pi l_{Bo}z_p^2 \alpha_p^2 c_o \tilde{U}_{pp}(q) \right\}\tilde{h}_p(q)\tilde{h}_p(-q)} \label{eq:heuristic_Icharged}
\end{eqnarray}
 where now (cf. Eq. ~\ref{eq:renorm_wpp})
\begin{eqnarray}
w_{pp,r} &=& w_{pp} + \frac{\Delta_p^2 (1-\alpha_p)^2}{c_o^2 \lambda_0}
\end{eqnarray}
Using $\tilde{M}(0) = 1/(8\pi \tilde{\kappa}_o(0)\tilde{\epsilon}^2(0))$, we get 
\begin{eqnarray}
I(0) &=& I_s(0)c_o V \left[\frac{1}{N} + w_{pp,r} c_o + 4\pi l_{Bo}z_p^2 \alpha_p^2 c_o \left[\frac{1}{\tilde{\kappa}_o^2(0)} -\frac{l_{Bo}z_p^2(1-\alpha_p)^2}{4\tilde{\kappa}_o(0)\tilde{\epsilon}^2(0)}\right]\right]^{-1}\label{eq:heuristic_Icharged_0}
\end{eqnarray}
which shows that screening effects tend to renormalize the excluded volume parameter similar 
to the local effects of the polarization. Furthermore, there is an additional contribution 
from the coupling between free and adsorbed counterions which tends to increase the scattering 
intensity at $q=0$ by decreasing the denominator in Eq. ~\ref{eq:heuristic_Icharged_0}. This additional term scales as $l_{Bo}^{3/2}/\tilde{\epsilon}^{2}(0)$. Further analysis of Eq. ~\ref{eq:heuristic_Icharged} reveals that the scattering 
intensity for charge regulating polymers can exhibit non-monotonicity. 
In addition to the non-monotonicity arising from non-zero values of $a_p$ and non-local effects of the polarization, there is an additional local maximum which appears due to the charge-charge correlations. In order to see the origin of this additional maximum, consider the limits of $a_p \rightarrow 0, \bar{\kappa}_o^2(q) \rightarrow 0, \alpha_p \rightarrow 1$ and optimize $I(0)/I(q)$ with respect to $q$ after using $g_D(x) \simeq 1/(1 + \beta_0 x), \beta_0 \simeq 0.5$. Optimization reveals that the additional maximum in $I(q)$ appear at 
$q = \left[24\pi l_{Bo}z_p^2 \alpha_p^2 c_o/(\beta_0 b^2)\right]^{1/4}$ due to interplay of 
chain connectivity and the \textit{purely repulsive} charge-charge Coulomb potential.   

Pairwise interactions in charge regulating polymers is fundamentally different and 
much more complicated than dipolar polymers.
In particular, 
Eq. ~\ref{eq:vq} reveals that an effective pairwise interactions in
charge regulating polymers are affected by the non-local dielectric effects (via $\tilde{\epsilon}(q)$), screening effects (via $\tilde{\kappa}_o^2(q)$) and an additional contribution ($\sim l_{Bo}^{3/2}$ as $\tilde{M}(0) \sim l_{Bo}^{-1/2}$) resulting from coupling of the adsorbed counterions on the chains with the ``free'' counterions. Screening effects and in particular, $\bar{\kappa}_o^2(0)$ has an additional contribution \textit{from the adsorbed counterions} and the non-local dielectric function
depends on the degree of ionization, $\alpha_p$. The dependence of $\bar{\kappa}_o^2(0)$ on $\alpha_p$ is in agreement with an expression for the screening length derived by Kirkwood and Shumaker\cite{kirkwood_forces}. Also, the effects of electrostatic fluctuations appear in the dielectric function via $\alpha_p$, which is also in agreement with another work by Kirkwood and Shumaker\cite{kirkwood_fluctuations}. Lastly, it can be readily shown that the non-local dielectric effects cause attractive interactions between the monomers and effective interactions between two monomers can be readily constructed to exhibit an attractive well along with oscillatory features. Both, the attraction and oscillatory nature of the effective interactions are hallmarks of electrostatics in finite sized particles\cite{kirkwood_stat,kirkwood_review,dressed_ions,muthu_double_screening}. 

In order to study effects of various parameters on the scattering intensity from charged polymers, a self-consistent calcualtion needs to be done by the mimization of the free energy with respect to $\alpha_p$ and construction of the free energy depends on specifics of
the charge regulation mechanism. We used a two-state model to construct the free energy and structure factor for charged polymers. \textcolor{black}{In the next section, we present a general theory which takes into account the effects of solvent molecules and added salts. Specific cases of salt-free ionic and dipolar polymer melts are considered by taking appropriate limits of the general theory.}

\section{Construction of a density functional theory}\label{sec:construct}
For studying the monomer density-density correlation function, we construct a monomer density functional 
theory after integrating out all other degrees of freedom. For such purposes, we use a field theory 
approach to decouple interactions and then consider perturbations about a homogeneous phase. For constructing 
the field theory, we consider the partition function for charge-regulating polyelectrolyte chains in the presence of finite-sized polar
solvent molecules, counterions and co-ions. The construction of the field theory is based on our previous work\cite{kumar_kilbey} \textcolor{black}{and a list of used symbols has been presented in the Supporting Information for the convenience of a reader}. 
For writing the partition function, we assume that there are $n_p$ mono-disperse (i.e., equal length) chains, each
containing $N$ Kuhn segments, each of length $b$. Following Edwards\cite{edwardsbook,fredbook}, each 
chain is represented as a continuous curve of length
$Nb$ so that $\mathbf{R}_{\alpha}(t_{\alpha})$ denotes the position vector for a particular segment, $t_{\alpha} \in (0,N)$,
along the backbone of $\alpha^{th}$ chain. Subscripts $p,s$ and $\gamma$ are used to
represent monomers, solvent molecules, small ions, respectively.
Three different kinds of small ions are considered and unless specified,
$\gamma = c, B^+$ and $A^-$ represents
counterions resulting from dissociation of the charged groups on the chains, cations and anions from the added salt, respectively.    
Here, we study negatively charged chains and the specificity of
the cations ($c$ and $B^+$) is taken into account to study the effects of different binding energies of the cations.
Generalization of the theoretical treatment here to the case of positively charged chains is straightforward. \textcolor{black}{Solvent molecules and ions are treated using the \textit{local} incompressiblity condition so that the total volume can be written as $V = n_p N/\rho_{po} + n_s/\rho_{so} + \sum_{\gamma}n_\gamma/\rho_{\gamma o}$ where $\rho_{po}$, $\rho_{so}$ and $\rho_{\gamma o}$ are
the number densities of the monomer, the solvent and the ions, respectively. We 
obtain results for compressible polymer melts by making appropriate substitutions related to size and dipole moment of the solvent molecules in the incompressible solution model. 
Furthermore, effects of ions in the incompressibility constraint are neglected at an appropriate place in this work to keep analytical calculations tractable and retained in the general model 
developed here to facilitate numerical work in future.}
$n_s$ and $n_{\gamma}$ are the total number of solvent molecules and small ions of type $\gamma$, respectively.
Also, $\mathbf{r}_k$ represents the
position vector of the $k^{th}$ small molecule like solvent molecules, counterions and coions.

Counterion adsorption on the polyelectrolyte chains is taken into account using
a two-state model, described in detail in our previous work\cite{kumar_kilbey}. Segments along the chains can be either in charged or in uncharged state.
To describe the two states, another arc length variable, $\theta_\alpha(t_\alpha)$, is introduced,
which enumerates the state of charging of the segment, $t_{\alpha}$ on $\alpha^{th}$ chain.
For the analysis here, $\theta_\alpha(t_\alpha) = 0$ means $t_\alpha$ is a neutral site
and $\theta_\alpha(t_\alpha) = 1$ represents a fully charged site along the backbone.
Like the average over all of the possible conformations in the theories of neutral polymers, a similar average over all of the possible charge distributions along the chains needs to be evaluated. We
represent the average over $\theta_\alpha(t_\alpha)$ by symbol
$\sum_{\left\{\theta_\alpha\right\}}\left < (\cdots) \right >$, which expicitly means
\begin{eqnarray}\label{eq:charge_sum}
\sum_{\left\{\theta_\alpha\right\}}\left < (\cdots) \right >  &=& \prod_{\alpha=1}^{n_p} \prod_{t_\alpha=0}^{N} \int D\left[\theta_\alpha(t_\alpha)\right](\cdots)p\left[\theta_\alpha(t_\alpha)\right]\Upsilon\left[\theta_\alpha(t_\alpha)\right].
\end{eqnarray}
Here, $p\left[\theta_\alpha\right]$ is the probability distribution function for the variable $\theta_\alpha$.
Also, $\Upsilon\left[\theta_\alpha(t_\alpha)\right]$ is the number of indistinguishable
ways in which $\theta_\alpha$ can be distributed among $n_pN$ sites for a fixed number of charged sites.
$\Upsilon\left[\theta_\alpha(t_\alpha)\right]$ takes into account the entropy of distribution of charged sites.

Like the segments, the counterions are also divided into two sets.
One set of counterions is ``free'' to explore the whole space and has translational
degrees of freedom. The other set is ``adsorbed'' on the backbone and
behave as electric dipoles (ion-pairs). The number of counterions in ``free'' and ``adsorbed''
states are taken to be $n_\gamma^f$ and $n_\gamma^a$, respectively, for
$\gamma = c, B^+$, so that $n_{\gamma} = n_\gamma^f + n_\gamma^a$. In the following, the dipole 
moment of a segment (in units of $e$, the charge of an electron) along the $\alpha^{th}$ chain backbone is
written as a vector, $\mathbf{p}_{\alpha}(t_\alpha) = p_p\mathbf{u}_{\alpha}(t_\alpha)$ so that 
 each dipole is of fixed length $p_p$ with its orientation depicted by $\mathbf{u}_{\alpha}(t_\alpha)$.
Similarly, $\mathbf{p}_{k}$ represents the dipole moment of the $k^{th}$ solvent
molecule with $p_s$ and $\mathbf{u}_{k}$ as its magnitude and orientation, respectively.

\textcolor{black}{
Electrostatic terms depend on the arc length variable $\theta_\alpha(t_\alpha)$.
This variable also determines the energetic contributions of counterion adsorption on the backbone, written as
$E\left\{\theta_\alpha\right\}$. Noting that
the dissociable groups on the chains have to dissociate first for the salt ions to
adsorb, it is written as
\begin{eqnarray}
E\left\{\theta_\alpha\right\} = (n_pN - n_{c}^a) \left[\mu_{p}^o + \mu_{c}^o - \mu_{pc}^o\right]
 + n_{B^+}^a \left[\mu_{pB}^o - \mu_{p}^o - \mu_{B^+}^o \right]\label{eq:binding_energy}
\end{eqnarray}
Here, $\mu_j^o$ is the chemical potential (in units of $k_B T$) for species of type $j$ in infinitely dilute conditions.
The differences in the chemical potentials are related to the equilibrium constants
of the corresponding reactions by the relations\cite{mcquarie}
\begin{eqnarray}
\mu_{p}^o + \mu_{c}^o - \mu_{pc}^o &=& 2.303 \mbox{pK}_{c} = -2.303 \log_{10}\mbox{K}_{c} \\
\mu_{p}^o + \mu_{B^+}^o - \mu_{pB}^o &=& 2.303 \mbox{pK}_{B^+} = -2.303 \log_{10}\mbox{K}_{B^+}
\end{eqnarray}
and we have defined $\mbox{K}_{c}$ and $\mbox{K}_{B^+}$ as the equilibrium (dissociation)
constants for the reactions
\begin{eqnarray}
pc &\rightleftharpoons& p^{-} + c^{+} \\
pB &\rightleftharpoons& p^- + B^+
\end{eqnarray}
respectively. Such a model of counterion adsorption was originally developed by Harris and Rice\cite{harris_rice}. We have used the same two state model in our previous works related to pH responsive polyelectrolyte brushes\cite{kumar_kilbey, mahalik_pe_brushes_2016}.
The probability distribution, $p$, needs to be determined self-consistently by the
minimization of the free energy and must
satisfy the relation $\int D\left[\theta_\alpha(t_\alpha)\right] p\left[\theta_\alpha(t_\alpha)\right] = 1$.
In this work, we take a variational \textit{ansatz} for $p$ and write it as
\begin{eqnarray}
p\left[\theta_\alpha(t_\alpha)\right] &=& \left(\sum_{\gamma'=c,B^+}\beta_{\gamma'}\right)
\delta\left[\theta_\alpha(t_\alpha)\right] + \left(1-\sum_{\gamma'=c,B^+}\beta_{\gamma'}\right)\delta\left[\theta_\alpha(t_\alpha) - 1\right] \label{eq:prob_dist}
\end{eqnarray}
so that $n_{\gamma'}^a = \beta_{\gamma'} n_p N$ for $\gamma' = c,B^+$.
Mathematically, $\beta_{c}$ and $\beta_{B^+}$ are the variational parameters, which will be determined by
minimization of the free energy. Physically, $\beta_{c}$ and $\beta_{B^+}$ correspond to
the fraction of sites on the chains occupied by $c$ and $B^+$, respectively.
\textcolor{black}{Treatment of $\beta_{c}$ and $\beta_{B^+}$ as variational parameters is equivalent to equating the electrochemical potential of the ions in ``free'' and ``adsorbed'' states\cite{hill_statmech}.} Furthermore, using Eq. ~\ref{eq:prob_dist} for the charge distribution
\begin{eqnarray} \label{eq:entropic_factor}
\Upsilon\left[\theta_\alpha(t_\alpha)\right] &\equiv& \Upsilon = \frac{(n_pN)!}{n_{c}^a! n_{B^+}^a!(n_pN- n_{c}^a - n_{B^+}^a)!}
\end{eqnarray}
Such a distribution is called ``annealed'' distribution in the literature\cite{borukhov}.
}

Using the notations described above, the
partition function ($Z$) for the polyelectrolyte chains can be written as\cite{kumar_kilbey}
\begin{eqnarray}
       Z & = & \int \prod_{\alpha=1}^{n_p} D[\mathbf{R}_\alpha] \sum_{\left\{\theta_\alpha(t_\alpha)\right\}}\left < \int \prod_{\alpha=1}^{n_p}\prod_{t_\alpha=0}^{N} d\mathbf{u}_{\alpha}(t_\alpha)
\int \frac{\Lambda^{-3n'}}{\prod_{\gamma'}n_{\gamma'}^f!n_{A^-}!n_p!n_s!}\prod_{\gamma}\prod_{j=1}^{n_{\gamma}} d\mathbf{r}_{j}
\right . \nonumber \\
&& \prod_{k=1}^{n_s}d\mathbf{r}_{k} \int \prod_{k=1}^{n_s}d\mathbf{u}_{k} \exp \left [-H_0\left\{\mathbf{R}_\alpha\right\}   - H_w\left\{\mathbf{R}_\alpha,\mathbf{r}_{k}\right\}
- H_{e}\left\{\mathbf{R}_\alpha,\mathbf{u}_{\alpha},\mathbf{r}_j, \mathbf{r}_k,\mathbf{u}_{k}\right\} -E\left\{\theta_\alpha\right\}\right ]\nonumber \\
&& \left . \prod_{\mathbf{r}}\mathbf{\delta}\left[\frac{\hat{\rho}_{p}(\mathbf{r})}{\rho_{po}} + \frac{\hat{\rho}_{s}(\mathbf{r})}{\rho_{so}} + \sum_{\gamma}\frac{\hat{\rho}_{\gamma}(\mathbf{r})}{\rho_{\gamma o}} - 1\right] \right > \label{eq:parti_sing}
\end{eqnarray}
where $\gamma' = c,B^+$, $\gamma = c,B^+, A^{-}$, $\Lambda$ is the de Broglie wavelength and 
$n' = n_p N + n_s + \sum_{\gamma} n_{\gamma}$. 

The Hamiltonian in Eq. ~\ref{eq:parti_sing} is written by taking into account
the contributions from the chain connectivity (given by $H_0$ in Eq. ~\ref{eq:connectivity} below),
the short ranged dispersion interactions (represented by $H_w$ in Eq. ~\ref{eq:dispersion})
and the long range electrostatic interactions
(written as $H_{e}$, which includes contributions from dipole-dipole, charge-dipole
and charge-charge interactions). For convenience in writing, in the following, we have
suppressed the explicit functional dependence of $H_{0},H_w$ and $H_{e}$.

Explicitly, contributions from the chain connectivity are given by:
\begin{eqnarray}
      H_0 &=& \frac {3}{2 b^2}\sum_{\alpha=1}^{n_p} \int_{0}^{N} dt_\alpha \left(\frac{\partial \mathbf{R}_\alpha(t_\alpha)}{\partial t_\alpha} \right )^{2} \label{eq:connectivity}
\end{eqnarray}
which represent flexible polymer chains\cite{edwardsbook}. Furthermore, $H_{w}$ 
takes into account the energetic contributions from short range dispersion interactions among different pairs. 
Following Edwards's formulation\cite{edwardsbook}, we model these interactions by
\begin{eqnarray}
H_w &=& \frac{1}{2}\int d\mathbf{r}
\left[w_{pp}\hat{\rho}_{p}^2(\mathbf{r}) + w_{ss}\hat{\rho}_{s}^2(\mathbf{r})  + 2 w_{ps} \hat{\rho}_{p}(\mathbf{r})
\hat{\rho}_{s}(\mathbf{r})\right] \label{eq:dispersion}
\end{eqnarray}
where, $w_{pp}, w_{ss}$ and $w_{ps}$ are the excluded volume parameters describing the
strength of interactions between $p-p, s-s$ and $p-s$ pairs, respectively. Also,
$\hat{\rho}_{p}(\mathbf{r})$ and $\hat{\rho}_{s}(\mathbf{r})$ represent the microscopic number
density of the monomers and the solvent molecules, respectively, at a
certain location $\mathbf{r}$ defined as
        \begin{eqnarray}
\hat{\rho}_{p}(\mathbf{r})  &=& \sum_{\alpha=1}^{n_p} \int_{0}^{N} dt_\alpha \, \hat{h}_p \left[\mathbf{r}-\mathbf{R}_\alpha(t_\alpha)\right] \\
\hat{\rho}_{s}(\mathbf{r})  &=& \sum_{k=1}^{n_s} \, \hat{h}_s \left[\mathbf{r}-\mathbf{r}_k\right]
\end{eqnarray}
where the functional form of $\hat{h}_j(\mathbf{r})$ characterizes the density distribution of a molecule of type $j$. It should be noted that in 
writing Eq. ~\ref{eq:dispersion}, we have ignored short-ranged interactions with counterions and coions. 

Electrostatic contributions to the Hamiltonian arising from charge-charge, charge-dipole and dipole-dipole interactions can be written
as (\textcolor{black}{see Supporting Information in \cite{kumar_kilbey}}) 
\begin{eqnarray}
H_e &=& \frac{l_{Bo}}{2}\int d\mathbf{r}\int d\mathbf{r}'
\frac{\left[\hat{\rho}_e(\mathbf{r}) - \nabla_{\mathbf{r}}.\hat{P}_{ave}(\mathbf{r}) \right]\left[\hat{\rho}_e(\mathbf{r}') - \nabla_{\mathbf{r}'}.\hat{P}_{ave}(\mathbf{r}')
\right]}{|\mathbf{r} - \mathbf{r}'|} \label{eq:final_particle_elec}
\end{eqnarray}
where $\hat{\rho}_e(\mathbf{r}) = \sum_{\gamma} z_\gamma \hat{\rho}_{\gamma}(\mathbf{r}) + z_p \hat{\rho}_{pe}(\mathbf{r})$
is the local charge density and $\hat{\rho}_\gamma(\mathbf{r})$ represents the local microscopic densities for the ions of type $\gamma$ at
$\mathbf{r}$, defined as
        \begin{eqnarray}
     \hat{\rho}_{\gamma}(\mathbf{r}) &=&  \sum_{j=1}^{n_{\gamma}} \hat{h}_{\gamma} \left[\mathbf{r}-\mathbf{r}_j\right] \quad \mbox{for} \quad \gamma = c,B^+,A^-.
\end{eqnarray}
Furthermore, $\hat{\rho}_{pe}(\mathbf{r})$ is the contribution to the charge density from the polyelectrolyte chains, given by
\begin{eqnarray}
     \hat{\rho}_{pe}(\mathbf{r}) &=& \sum_{\alpha=1}^{n_p} \int_{0}^{N} dt_\alpha \, \hat{h}_p \left[\mathbf{r} -\mathbf{R}_\alpha(t_\alpha)\right] \theta_\alpha(t_\alpha)
\end{eqnarray}
and 
$\hat{P}_{ave}(\mathbf{r}')
= \int d\mathbf{u} \hat{P}(\mathbf{r},\mathbf{u})\mathbf{u}$, is an angularly averaged polarization so that
$\hat{P}(\mathbf{r},\mathbf{u} ) = p_p \bar{\rho}_p(\mathbf{r},\mathbf{u} ) + p_s \bar{\rho}_s(\mathbf{r},\mathbf{u} )$
is local polarization density at $\mathbf{r}$ in a direction specified by $\mathbf{u}$. Formally, 
\begin{eqnarray}
    \bar{\rho}_{p}(\mathbf{r},\mathbf{u})  &=& \sum_{\alpha=1}^{n_p} \int_{0}^{N} dt_\alpha \, \hat{h}_p \left[\mathbf{r}
-\mathbf{R}_\alpha(t_\alpha)\right]
\delta \left[\mathbf{u}-\mathbf{u}_\alpha(t_\alpha)\right]\left[1 - \theta_\alpha(t_\alpha)\right] \\
\bar{\rho}_{s}(\mathbf{r},\mathbf{u})  &=& \sum_{k=1}^{n_s} \, \hat{h}_s \left[\mathbf{r}-\mathbf{r}_k\right]
\delta \left[\mathbf{u}-\mathbf{u}_k\right]
\end{eqnarray}

Using Eq. ~\ref{eq:prob_dist}, we can write Eq. ~\ref{eq:parti_sing} in a field theoretic form (see Appendix A for the details)
\begin{eqnarray}
       Z & = & e^{-F_a/k_B T}\int \prod_{\alpha=1}^{n_p} D[\mathbf{R}_\alpha]\int \prod_{k=1}^{n_{s}} d\mathbf{r}_{k} \exp \left [-H_0\left\{\mathbf{R}_\alpha\right\}
- H_w\left\{\mathbf{R}_\alpha,\mathbf{r}_{k}\right \} -H_{elec}\left\{\mathbf{R}_\alpha,\mathbf{r}_{k}\right \}\right ] \nonumber \\
&& \label{eq:parti_elec_integrated}
\end{eqnarray}
where
\begin{eqnarray}
       \exp \left [-H_{elec}\left\{\mathbf{R}_\alpha,\mathbf{r}_{k}\right \}\right ]  & = & \int \frac{D\left[\psi\right]}{\zeta_\psi}\int D\left[\eta\right] \exp \left [- \frac{\bar{H}_{elec}}{k_BT}\right ] \label{eq:parti_abstract}
\end{eqnarray}
and $\bar{H}_{elec}$ is given by
\begin{eqnarray}
       \frac{\bar{H}_{elec}}{k_BT} &=& - \frac{1}{8\pi l_{Bo}}\int d\mathbf{r} \psi(\mathbf{r})\nabla_{\mathbf{r}}^2 \psi(\mathbf{r}) 
+ i \int d\mathbf{r} \hat{\rho}_p(\mathbf{r})\psi_p(\mathbf{r}) \nonumber \\
&& - \sum_{\gamma' = c,B^{+}}n_{\gamma'}^f \ln \bar{Q}_{\gamma'}\left\{\psi,\eta\right\} - n_{A^-}\ln \bar{Q}_{A^-}\left\{\psi,\eta\right\} + i \int d\mathbf{r}\eta(\mathbf{r})\left[\sum_{j=p,s}\frac{\hat{\rho}_j(\mathbf{r})}{\rho_{j0}}-1\right] \nonumber \\
&&- \int d\mathbf{r}'\hat{\phi}_s(\mathbf{r}')\ln \left[\frac{\sin\left\{p_s \left|\int d\mathbf{r} \psi(\mathbf{r})\nabla_\mathbf{r}\hat{h}_s(\mathbf{r}-\mathbf{r}')\right|\right\}}
{p_s \left|\int d\mathbf{r} \psi(\mathbf{r})\nabla_\mathbf{r}\hat{h}_s(\mathbf{r}-\mathbf{r}')\right|}\right] \label{eq:hami_physical3}
\end{eqnarray}
and $\bar{Q}_{\gamma}$ is the partition function for an ion of type $\gamma$, given by
\begin{eqnarray}
\bar{Q}_{\gamma = c,B^{+},A^{-}}\left\{\psi,\eta\right\} &=& \int d \mathbf{r} \exp\left[- i \int d\mathbf{r}' \hat{h}_{\gamma}(\mathbf{r}-\mathbf{r}')\left\{z_{\gamma}\psi(\mathbf{r}') + \eta(\mathbf{r}')/\rho_{\gamma0}\right\}\right]
\end{eqnarray}

Also, 
\begin{eqnarray}
       \frac{F_a}{k_B T} & = & n_{B^+}^a\ln \mbox{K}_{B^+} - (n_pN-n_{c}^a)\ln \mbox{K}_{c}
- \ln \left[\frac{n_pN!}{n_{c}^a! n_{B^+}^a!(n_pN- n_{c}^a - n_{B^+}^a)!}\right] \nonumber \\
&& + \ln \left[n_{c}^f! n_{B^+}^f! n_{A^-}!n_s!n_p!\right] - (n_pN + n_s) \ln 4\pi + 3n'\ln \Lambda
\end{eqnarray}
Furthermore, we have defined, $\hat{\phi}_s(\mathbf{r})$ and $\psi_p(\mathbf{r})$ by the relations
\begin{eqnarray}
       \hat{\rho}_s(\mathbf{r}) &=& \int d\mathbf{r}' \hat{h}_s(\mathbf{r}-\mathbf{r}') \hat{\phi}_s(\mathbf{r}') \\
i \int d\mathbf{r} \hat{\rho}_p(\mathbf{r})\psi_p(\mathbf{r}) &=& -\sum_{\alpha=1}^{n_p}\int_0^N dt_{\alpha} \ln \left[(1 - \beta_{c} - \beta_{B^+}) \exp\left[-i z_p \int d \mathbf{r} 
\hat{h}_p(\mathbf{r}-\mathbf{R}_{\alpha})\psi(\mathbf{r})\right]
\right . \nonumber \\
&& \left . + \left(\sum_{\gamma = c,B^+} \beta_{\gamma}\right) \left[\frac{\sin\left(p_p \left|\int d\mathbf{r}\psi(\mathbf{r})
\nabla_{\mathbf{r}}\hat{h}_{p}(\mathbf{r}-\mathbf{R}_{\alpha})\right|\right)}
{p_p \left|\int d\mathbf{r}\psi(\mathbf{r})\nabla_{\mathbf{r}}\hat{h}_
{p}(\mathbf{r}-\mathbf{R}_{\alpha})\right|} \right]\right]
\end{eqnarray}

  \subsection{Electrostatics of a weakly inhomogeneous phase}
For studing a weakly inhomogeneous phase, we consider perturbations of densities and electrostatic potential about a homogeneous phase. In particular, we write $i\psi(\mathbf{r}) = \psi_b + i\delta \psi(\mathbf{r}), \hat{\phi}_p(\mathbf{r}) = n_pN/V + \delta \hat{\phi}_p(\mathbf{r}),
\hat{\phi}_s(\mathbf{r}) = n_s/V + \delta \hat{\phi}_s(\mathbf{r})$ so that $\int d\mathbf{r} \delta \psi(\mathbf{r}) = \int d\mathbf{r} \delta \hat{\phi}_p(\mathbf{r}) = \int d\mathbf{r}\delta \hat{\phi}_s(\mathbf{r}) = 0$. \textcolor{black}{In expanding $\psi(\mathbf{r})$, we have used the fact that the saddle-point for $\psi(\mathbf{r})$, which represents a homogeneous phase, lies along the imaginary axis in the complex plane}. Furthermore, \textcolor{black}{we assume that $\int d\mathbf{r}' \hat{h}_{\gamma}(\mathbf{r}-\mathbf{r}')\eta(\mathbf{r}')/\rho_{\gamma0} \rightarrow 0$ i.e., we neglect effects of the finite size of the ions on the local 
incompressibility constraint so that $n_p N + n_s = V$. Such an assumption is valid in dilute limit of small ions such as in the case of salt-free solutions and melts\cite{kirkwood_stat,kirkwood_review}. However, effects of finite size of the ions in the incompressibility constraint can be included in numerical calculations and we ignore such effects here to keep analytical calculations tractable.} For a weakly inhomogeneous phase, we can write $\bar{H}_{elec}$ as (cf. Eq. ~\ref{eq:hami_physical3})
\begin{eqnarray}
\frac{\bar{H}_{elec}}{k_BT} &=& \frac{1}{8\pi l_{Bo}}\int d\mathbf{r} \int d\mathbf{r}'\delta \psi(\mathbf{r})\bar{G}_0(\mathbf{r},\mathbf{r}') \delta \psi(\mathbf{r}') + \frac{z_p^2 w_{cr}}{2}\int d\mathbf{r} \int d\mathbf{r}'\delta \psi(\mathbf{r})\bar{G}_1(\mathbf{r},\mathbf{r}') \delta \psi(\mathbf{r}') \nonumber \\
&& + i z_p \alpha_p \int d\mathbf{r} \hat{\rho}_p(\mathbf{r})\delta \psi(\mathbf{r})
+ i \int d\mathbf{r}\eta(\mathbf{r})\left[\sum_{j=p,s}\frac{\delta \hat{\rho}_j(\mathbf{r})}{\rho_{j0}}\right] \nonumber \\
&& + n_p N \psi_{p,b} + \psi_b \left[\sum_{\gamma' = c,B^+}z_{\gamma'}n_{\gamma'}^f + z_{A^-}n_{A^-}\right]- \left[\sum_{\gamma' = c,B^{+}}n_{\gamma'}^f + n_{A^-}\right]\ln V\label{eq:perturb1}
\end{eqnarray}
where
\begin{eqnarray}
\bar{G}_0(\mathbf{r}, \mathbf{r}^{'})  & = & - \nabla_{\mathbf{r}}^2 \delta(\mathbf{r}-\mathbf{r}') 
+ \frac{4\pi l_{Bo}p_s^2} {3}
\int d\mathbf{r}'' \hat{\phi}_{s}(\mathbf{r}'')\nabla_{\mathbf{r}}\hat{h}_s(\mathbf{r}-\mathbf{r}'').\nabla_{\mathbf{r}'}\hat{h}_s(\mathbf{r}'-\mathbf{r}'') \nonumber \\
&& + \frac{4\pi l_{Bo}} {3}\sum_{\gamma' = c,B^+}\left[\frac{\beta_{\gamma'}p_p^2}{e^{-\psi_{p,b}}}\right]
\int d\mathbf{r}'' \hat{\phi}_{p}(\mathbf{r}'')\nabla_{\mathbf{r}}\hat{h}_p(\mathbf{r}-\mathbf{r}'').\nabla_{\mathbf{r}'}\hat{h}_p(\mathbf{r}'-\mathbf{r}'') \nonumber \\
&& + \frac{4\pi l_{Bo}}{V}\int d\mathbf{r}''\left[\sum_{\gamma'=c,B^+} z_{\gamma'}^2 n_{\gamma'}^f \hat{h}_{\gamma'}(\mathbf{r}-\mathbf{r}'')\hat{h}_{\gamma'}(\mathbf{r}'-\mathbf{r}'')\right.\nonumber \\
&& \left . + z_{A^-}^2 n_{A^-} \hat{h}_{A^-}(\mathbf{r}-\mathbf{r}'')\hat{h}_{A^-}(\mathbf{r}'-\mathbf{r}'') 
 + z_p^2 w_{cr} n_p N \hat{h}_{p}(\mathbf{r}-\mathbf{r}'')\hat{h}_{p}(\mathbf{r}'-\mathbf{r}'') \right]\label{eq:operator0}
\end{eqnarray}
and 
\begin{eqnarray}
\bar{G}_1(\mathbf{r}, \mathbf{r}^{'})  & = & \int d\mathbf{r}'' \delta \hat{\phi}_{p}(\mathbf{r}'')\hat{h}_p(\mathbf{r}-\mathbf{r}'')
\hat{h}_p(\mathbf{r}'-\mathbf{r}'') \label{eq:operator1}
\end{eqnarray}
Here, we have defined parameters characterizing electrostatics in the homogeneous phase as 
\begin{eqnarray}
\alpha_p &=& \frac{1-\sum_{\gamma'=c,B^+}\beta_{\gamma'}}{\exp\left[-\psi_{p,b}\right]}\exp\left[-z_p\psi_{b}\right] \label{eq:para1}\\
w_{cr} &=& \frac{\alpha_p \sum_{\gamma'=c,B^+}\beta_{\gamma'}}{\exp\left[-\psi_{p,b}\right]} \label{eq:para2}\\
\exp\left[-\psi_{p,b}\right] &=& \left(\sum_{\gamma'=c,B^+}\beta_{\gamma'}\right) + \left(1-\sum_{\gamma'=c,B^+}\beta_{\gamma'}\right) \exp\left[-z_p\psi_{b}\right] \label{eq:para3}
\end{eqnarray}
Also, $\hat{\rho}_p(\mathbf{r}) = \int d\mathbf{r}' \hat{h}_p(\mathbf{r}-\mathbf{r}')\hat{\phi}_{p}(\mathbf{r}') = n_p N + \delta \hat{\rho}_p(\mathbf{r}), \hat{\rho}_s(\mathbf{r}) = \int d\mathbf{r}' \hat{h}_s(\mathbf{r}-\mathbf{r}')\hat{\phi}_{s}(\mathbf{r}') = n_s + \delta \hat{\rho}_s(\mathbf{r})$. 
Plugging Eq. ~\ref{eq:perturb1} in Eq. ~\ref{eq:parti_abstract} and expanding in powers of $z_p^2 w_{cr}$, we get
\begin{eqnarray}
 H_{elec}\left\{\mathbf{R}_\alpha,\mathbf{r}_{k}\right \}  \simeq \frac{F_o}{k_BT} + \frac{z_p^2 w_{cr}}{2}\int d\mathbf{r}_1 \int d\mathbf{r}_2 \left<\delta \psi(\mathbf{r}_1)\delta \psi(\mathbf{r}_2)\right> \bar{G}_1(\mathbf{r}_1,\mathbf{r}_2) && \nonumber \\
 - \frac{z_p^4 w_{cr}^2}{8}\int d\mathbf{r}_1 \int d\mathbf{r}_2 \int d\mathbf{r}_3 \int d\mathbf{r}_4\left<\delta \psi(\mathbf{r}_1)\delta \psi(\mathbf{r}_2) \delta \psi(\mathbf{r}_3)\delta \psi(\mathbf{r}_4)\right> \bar{G}_1(\mathbf{r}_1,\mathbf{r}_2)\bar{G}_1(\mathbf{r}_3,\mathbf{r}_4) && \nonumber \\
 + \frac{4\pi l_{Bo} z_p^2 \alpha_p^2}{2}\int d\mathbf{r}_1 \int d\mathbf{r}_2 \hat{\rho}_p(\mathbf{r}_1)\bar{G}_0^{-1}(\mathbf{r}_1, \mathbf{r}_{2}) \hat{\rho}_p(\mathbf{r}_2)- \ln \left[\prod_{\mathbf{r}}\mathbf{\delta}\left[\sum_{j=p,s}\frac{\delta\hat{\rho}_{j}(\mathbf{r})}{\rho_{j0}}\right]\right] && \nonumber \\
 -\ln \left[\frac{D[\psi]}{\zeta_{\psi}}\exp\left[-\frac{1}{8\pi l_{Bo}}\int d\mathbf{r} \int d\mathbf{r}'\delta \psi(\mathbf{r})\bar{G}_0(\mathbf{r},\mathbf{r}') \delta \psi(\mathbf{r}')\right]\right] && \label{eq:perturb2}
\end{eqnarray}
so that 
\begin{eqnarray}
\frac{F_o}{k_BT} &=& n_p N \psi_{p,b} + \psi_b \left[\sum_{\gamma' = c,B^+}z_{\gamma'}n_{\gamma'}^f + z_{A^-}n_{A^-}\right]- \left[\sum_{\gamma' = c,B^{+}}n_{\gamma'}^f + n_{A^-}\right]\ln V
\end{eqnarray}

Here, we have defined $\bar{G}_0^{-1}(\mathbf{r}_1,\mathbf{r}_2)$ using the relation 
\begin{eqnarray}
\int d\mathbf{r}_3 \bar{G}_0(\mathbf{r}_1,\mathbf{r}_3) \bar{G}_0^{-1}(\mathbf{r}_3,\mathbf{r}_2) & = & \delta (\mathbf{r}_1 - \mathbf{r}_2)\label{eq:inv_formal}
\end{eqnarray}
and $<\cdots>$ means
\begin{eqnarray}
\left<(\cdots)\right> &=& \frac{\int D[\psi](\cdots)\exp\left[-\frac{1}{8\pi l_{Bo}}\int d\mathbf{r} \int d\mathbf{r}'\delta \psi(\mathbf{r})\bar{G}_0(\mathbf{r},\mathbf{r}') \delta \psi(\mathbf{r}')\right]}{\int D[\psi]\exp\left[-\frac{1}{8\pi l_{Bo}}\int d\mathbf{r} \int d\mathbf{r}'\delta \psi(\mathbf{r})\bar{G}_0(\mathbf{r},\mathbf{r}') \delta \psi(\mathbf{r}')\right]}\label{eq:avg_def}
\end{eqnarray} 
\textcolor{black}{All of the averages appearing in Eq. ~\ref{eq:perturb2} are over a probability distribution, which is Gaussian, and can be computed leading to (see the Supporting Information for the details) 
\begin{eqnarray}
Z\left\{\delta \hat{\rho}_p\right\} &=& V^{n_s} e^{-F_a/k_B T - H_{w0}}\zeta_w\int \prod_{\alpha=1}^{n_p} D[\mathbf{R}_\alpha]
\exp \left [-H_0\left\{\mathbf{R}_\alpha\right\}
+  \bar{\chi}_{ps}\int d\mathbf{r} \left[ \frac{\delta \hat{\rho}_p(\mathbf{r})}{\rho_{p0}}\right]^2 \right . \nonumber \\
&& \left . - \frac{\rho_{s0}^2}{2\rho_{p0}^2} \int d\mathbf{r} \int d\mathbf{r}' \delta \hat{\rho}_p(\mathbf{r}) J^{-1}(\mathbf{r}-\mathbf{r}')\delta \hat{\rho}_p(\mathbf{r}')
- \bar{F}_{elec}\left\{\mathbf{R}_\alpha\right \}\right ] \label{eq:parti_poly}
\end{eqnarray}
so that
\begin{eqnarray}
H_{w0} &=& \frac{1}{2}\left[w_{pp}n_p N \rho_{p0} + w_{ss}n_s\rho_{s0}\right] + \bar{\chi}_{ps} V \frac{n_p N/V}{\rho_{p0}}\frac{n_{s}/V}{\rho_{s0}}
\end{eqnarray}
where $\bar{\chi}_{ps}$ is defined by
\begin{eqnarray}\bar{\chi}_{ps} &=& w_{ps}\rho_{p0}\rho_{s0} - \frac{w_{pp}\rho_{p0}^2 + w_{ss}\rho_{s0}^2}{2} \label{eq:chi_parameter1}
\end{eqnarray}
Furthermore, 
\begin{eqnarray}
 \bar{F}_{elec}\left\{\mathbf{R}_\alpha\right \}  & = & \frac{F_o}{k_BT} - \frac{(4\pi l_{Bo})^2 z_p^4 w_{cr}^2}{4}\sum_{\mathbf{q}\neq 0}\tilde{\delta \phi}_p(\mathbf{q})\tilde{\delta \phi}_p(-\mathbf{q})\tilde{U}(\mathbf{q}) \nonumber \\
&& + \frac{4\pi l_{Bo} z_p^2 \alpha_p^2 V}{2}\sum_{\mathbf{q}} \frac{\tilde{\rho}_p(\mathbf{q})\tilde{\rho}_p(-\mathbf{q})}{m(\mathbf{q})}
+ \frac{1}{2}\sum_{\mathbf{q}\neq 0}\ln \left[\frac{m(\mathbf{q})}{\mathbf{q}^2} \right]\label{eq:perturb3_redef}\\
\tilde{U}(\mathbf{q}) &=& \sum_{\mathbf{q}_1} \frac{\tilde{h}_p(\mathbf{q}_1)\tilde{h}_p(-\mathbf{q}_1)\tilde{h}_p(\mathbf{q}_1-\mathbf{q})\tilde{h}_p(\mathbf{q}-\mathbf{q}_1)}{m(\mathbf{q}_1)m(\mathbf{q}-\mathbf{q}_1)}\label{eq:capM}\\
m(\mathbf{q}_1) &=& \mathbf{q}_1^2 (1 + \Delta\tilde{\epsilon_h}(\mathbf{q}_1)) + \tilde{\kappa}^2(\mathbf{q}_1) + \sum_{\mathbf{q}_3}\tilde{g}(\mathbf{q}_3,-\mathbf{q}_1,\mathbf{q}_3-\mathbf{q}_1)(\mathbf{q}_1\cdot \mathbf{q}_3) \label{eq:mfactor}\\
\Delta\tilde{\epsilon_h}(\mathbf{q}_1)) &=& \frac{4\pi l_{Bo}p_s^2 n_s}{3V}\tilde{h}_s(\mathbf{q}_1)\tilde{h}_s(-\mathbf{q}_1) + \frac{4\pi l_{Bo}p_p^2 n_p N}{3V}\left[\sum_{\gamma' = c,B^+}\frac{\beta_{\gamma'}}{e^{-\psi_{p,b}}}\right]\tilde{h}_p(\mathbf{q}_1)\tilde{h}_p(-\mathbf{q}_1)\label{eq:homo_ft}\\
\tilde{\kappa}^2(\mathbf{q}_1) &=& 4\pi l_{Bo} \sum_{\gamma = c,B^+,A^-,p}z_\gamma^2 \Gamma_\gamma \tilde{h}_\gamma(\mathbf{q}_1)\tilde{h}_\gamma(-\mathbf{q}_1)\label{eq:kappa_ft}\\
\tilde{g}(\mathbf{q}_1,\mathbf{q}_2,\mathbf{q}_3) &=& \frac{4\pi l_{Bo}p_s^2}{3}
\tilde{\delta \phi}_{s}(\mathbf{q}_3)\tilde{h}_s(\mathbf{q}_1)\tilde{h}_s(\mathbf{q}_2) + \frac{4\pi l_{Bo}p_p^2}{3}\left[\sum_{\gamma' = c,B^+}\frac{\beta_{\gamma'}}{e^{-\psi_{p,b}}}\right]
\tilde{\delta \phi}_{p}(\mathbf{q}_3)\tilde{h}_p(\mathbf{q}_1)\tilde{h}_p(\mathbf{q}_2)\nonumber \\
&& \label{eq:inhomo_ft}
\end{eqnarray}
where, again, $\mathbf{q}$ is a wave-vector and all of the quantities in the Fourier space are represented by the superscript \textasciitilde. Use of the convolution theorem and local incompressibility constraint leads to $\tilde{\rho}_{j}(\mathbf{q}) = \tilde{h}_{j}(\mathbf{q})\tilde{\phi}_j(\mathbf{q})$ and
$\tilde {\delta\phi}_s(\mathbf{q}) = -\rho_{s0}\tilde {\delta\phi}_p(\mathbf{q})\tilde {h}_p(\mathbf{q})/(\rho_{p0}\tilde {h}_s(\mathbf{q}))$, respectively. Also, we have defined 
$\Gamma_p = w_{cr}n_p N/V, \Gamma_{\gamma' = c,B^+} = n_{\gamma'}^f/V, \Gamma_{A^-} = n_{A^-}/V$.
Integration over tranlational degree of freedom of the solvent leads to additional contributions in the partition function (see the Supporting Information for the details), given by
\begin{eqnarray}
\zeta_w &=& \int D\left[w_s\right]\exp \left [- \frac{1}{2} \int d\mathbf{r} \int d\mathbf{r}'w_s(\mathbf{r})J(\mathbf{r}-\mathbf{r}')w_s(\mathbf{r}')\right ] \\
J(\mathbf{r}-\mathbf{r}') &=& \frac{n_s}{V}\int d\mathbf{r}_1 \hat{h}_s(\mathbf{r}-\mathbf{r}_1)\hat{h}_s(\mathbf{r}_1-\mathbf{r}') \\
\int d\mathbf{r}' J(\mathbf{r}-\mathbf{r}')J^{-1}(\mathbf{r}'-\mathbf{r}'') &=& \delta(\mathbf{r}-\mathbf{r}'') 
\end{eqnarray}
}
Eq. ~\ref{eq:parti_poly} is the desired effective one-component description of the charge regulating polymer solutions. In the next section, we use this to construct a density functional theory for weakly inhomogeneous phases. 
 
\section{Results} \label{sec:results}
Here, we use Eq. ~\ref{eq:parti_poly} to construct the free energy of a homogeneous phase and weakly inhomogeneous phases. Two special cases of weakly inhomogeneous phases are 
considered. In one case, we assume that all of counterions are bound on the chains. This particular case is relevant for studying dipolar polymers in an electrolyte solution. In another case, we consider weakly 
inhomogeneous phase containing partially neutralized polymers in an electrolyte solution. Numerical 
evaluations of the density-density correlation function for the partially neutralized polymers are done by minimizing the free energy of a salt-free weakly inhomogeneous phase with respect to the parameter $\alpha_p$, which represents the average degree of ionization of the chains. Comparisons with small angle X-ray scattering experiments on salt-free dipolar and ionic polymer melts are presented.  
 
  \subsection{Homogeneous phase}\label{app:F}
For a homogeneous phase, $\tilde{\delta\phi}_p(\mathbf{q}) = 0$. In this case, Eq. ~\ref{eq:parti_poly} gives the free energy of the homogeneous phase, 
written as ($F_h/k_B T = -\ln Z\left\{\delta \hat{\rho}_p = 0\right\}$)
\begin{eqnarray}
\frac{F_h}{k_B T} &=& \frac{F_a}{k_B T} + \frac{F_o}{k_B T} - n_s\ln V +  H_{w0} - \ln \zeta_w -n_p \ln Q_p\{0\}
\nonumber \\
&&  + \frac{4\pi l_{Bo} z_p^2 \alpha_p^2 V}{2}\frac{\left[n_p N/V\right]^2}{\tilde{\kappa}^2(0)}
+ \frac{1}{2}\sum_{\mathbf{q}\neq 0}\ln \left[1 + \Delta \tilde{\epsilon}_h(\mathbf{q}) + \frac{\tilde{\kappa}^2(\mathbf{q})}{\mathbf{q}^2} \right]\label{eq:free_homo}
\end{eqnarray}
where
\begin{eqnarray}
Q_p\{0\} &=& \int D[\mathbf{R}]\exp \left [-\frac {3}{2 b^2}\int_{0}^{N} dt \left(\frac{\partial \mathbf{R}(t)}{\partial t} \right )^{2}\right]
\end{eqnarray}
It is worthwhile to consider the case of spherically symmetric molecules so that $\hat{h}_j(\mathbf{r}) = \exp\left[-\pi r^2/2a_j^2\right]/(2a_j^2)^{3/2}$ for $j=p,s,c,B^+,A^-$ 
with $a_j = a$. Using the notation $r = |\mathbf{r}|, 
q = |\mathbf{q}|$, for the case of spherically symmetric molecules $\tilde{h}_j(\mathbf{q}) = \tilde{h}(q) = \exp\left[-q^2a^2/2\pi\right]$. In the continuum limit, $\sum_{\mathbf{q}} \equiv V\int d\mathbf{q}/(2\pi)^3$, 
\begin{eqnarray}
\frac{1}{2}\sum_{\mathbf{q}\neq 0}\ln \left[1 + \Delta \tilde{\epsilon}_h(\mathbf{q}) + \frac{\tilde{\kappa}^2(\mathbf{q})}{\mathbf{q}^2} \right] 
= \frac{1}{2}\sum_{\mathbf{q}\neq 0}\left\{\ln \left[1 + \Delta \tilde{\epsilon}_h(\mathbf{q})\right] + \ln \left[1 + \frac{\tilde{\kappa}^2(\mathbf{q})}{\mathbf{q}^2\left[1 + \Delta \tilde{\epsilon}_h(\mathbf{q})\right]} \right]\right\}&&\nonumber \\
=-\frac{V}{16 a^3}L_{5/2}(2-\epsilon_s -\epsilon_p))
+ V\left[\frac{\kappa_h^2}{8\sqrt{\pi}a}-\frac{\kappa_h^3}{12\pi} + \frac{3}{32\pi^{3/2}}\kappa_h^4 a + O(a^2)\right]&&\nonumber \\
&&\label{eq:free_homo_symm}
\end{eqnarray}
where both the integrals can be calculated exactly for the case of spherically symmetric molecules. The first integral is shown exactly as a series and the asymptotic limit of the second 
integral in the series of $a$ are presented. Here, we have defined 
\begin{eqnarray}
\epsilon_s &=& 1 + \frac{4\pi l_{Bo}}{3}p_s^2 \frac{n_s}{V}\\
\epsilon_p &=& 1 + \frac{4\pi l_{Bo}}{3}p_p^2 \frac{n_p N}{V}\left[\sum_{\gamma'=c,B^+}\frac{\beta_{\gamma'}}{e^{-\psi_{p,b}}}\right]\\
\kappa_h^2 &=& 4\pi l_{B}\left[\sum_{\gamma=c,B^+,A^-,p}z_\gamma^2\Gamma_\gamma \right] \label{eq:screen_homo}\\
l_{B}  &=& \frac{l_{Bo}}{1 + \frac{4\pi l_{Bo}}{3}p_s^2 \frac{n_s}{V} + + \frac{4\pi l_{Bo}}{3}p_p^2 \frac{n_p N}{V}\left[\sum_{\gamma'=c,B^+}\frac{\beta_{\gamma'}}{e^{-\psi_{p,b}}}\right]}\\
L_{k}(x) &=& \sum_{s=1}^{\infty}\frac{x^s}{s^k}
\end{eqnarray} 
so that $L_{k}(x)$ is the Lambert-W function of fractional order\cite{Corless1996}. 

Electrostatic contributions to the homogeneous phase, given by Eq. ~\ref{eq:free_homo_symm} 
has two contributions. The first contribution is related to the fluctuations in polarization and depends on the dielectric constants ($\epsilon_s$ and $\epsilon_p$ for homogeneous solvent and polymer, respectively) and length scale, $a$, of the smeared distribution. It should be noted that the dielectric constants $\epsilon_s$ and $\epsilon_p$ are the same as predicted by the Langevin-Debye model\cite{dielectric}. The second electrostatic contribution to the free 
energy of a homogeneous phase results has the ionic self-energy part ($\sim 1/a$), the Debye-H\"{u}ckel correlation energy ($-\kappa_h^3$) and higher order terms in powers of $a$. 
However, the screening length $\kappa_h^{-1}$ (cf. Eq. ~\ref{eq:screen_homo}) depends on the concentration of all charged species including the counterions which are adsorbed on the chains. Furthermore, dependence of 
$\kappa_h^{-1}$ on $\alpha_p$ (the degree of ionization) leads to an implicit dependence of 
the screening length on number density of monomers, $n_p N/V$.
   
  \subsection{Weakly inhomogeneous phase}\label{app:G}
For constructing free energy of a weakly inhomogeneous phase and monomer density-density correlation function, we rewrite Eq. ~\ref{eq:parti_poly} in a form
\begin{eqnarray}
Z\left\{\delta \hat{\rho}_p\right\} &=& e^{-F_h/k_B T}\frac{1}{Q_p\{0\}^{n_p}}\int \prod_{\alpha=1}^{n_p} D[\mathbf{R}_\alpha]
\exp \left [-H_0\left\{\mathbf{R}_\alpha\right\} - H_{int}\left\{\mathbf{R}_\alpha\right\}\right]\label{eq:parti_inhomo}
\end{eqnarray}
where
\begin{eqnarray}
H_{int}\left\{\mathbf{R}_\alpha\right\} &=&
  \frac{V}{2\rho_{p0}^2}\sum_{\mathbf{q}\neq 0}\left[\rho_{s0}^2\tilde{J}^{-1}(\mathbf{q})-2\bar{\chi}_{ps}\right]\tilde{\delta \rho}_p(\mathbf{q})\tilde{\delta \rho}_p(-\mathbf{q}) + \frac{4\pi l_{Bo} z_p^2 \alpha_p^2 V}{2}\sum_{\mathbf{q}\neq 0} \frac{\tilde{\delta \rho}_p(\mathbf{q})\tilde{\delta \rho}_p(-\mathbf{q})}{m(\mathbf{q})} \nonumber \\
&& - \frac{(4\pi l_{Bo})^2 z_p^4 w_{cr}^2}{4}\sum_{\mathbf{q}\neq 0}\tilde{\delta \phi}_p(\mathbf{q})\tilde{\delta \phi}_p(-\mathbf{q})\tilde{U}(\mathbf{q}) 
+ \frac{1}{2}\sum_{\mathbf{q}\neq 0}\ln \left[\frac{m(\mathbf{q})}{\mathbf{q}^2 \left[1 + \Delta \tilde{\epsilon}_h(\mathbf{q})\right] + \tilde{\kappa}^2(\mathbf{q})} \right]
\nonumber \\
&&
\end{eqnarray}
where $\tilde{\delta \rho}_p(\mathbf{q}) = \tilde{h}_p(\mathbf{q})\tilde{\delta \phi}_p(\mathbf{q})$ and $\tilde{J}^{-1}(\mathbf{q}) = 1/\tilde{J}(\mathbf{q}) = V/(n_s \tilde{h}_s(\mathbf{q})\tilde{h}_s(-\mathbf{q}))$. If $\tilde{h}_p(\mathbf{q})$ is an even function of $\mathbf{q}$ and a real number, then we have the relations $\tilde{\delta \rho}_p^{\star}(\mathbf{q})= \tilde{\delta \rho}_p(-\mathbf{q}), \tilde{\delta \phi}_p^{\star}(\mathbf{q})= \tilde{\delta \phi}_p(-\mathbf{q})$, where the superscript $\star$ means the complex conjugate. Introducing a collective density ($\tilde{c}(\mathbf{q})$) and field variables in the Fourier space\cite{edwardsbook} for $\tilde{\delta \phi}_p(\mathbf{q})$,
expanding the chain partition function in powers of the field variables up to quadratic terms and integrating out the field variables, we can rewrite Eq. ~\ref{eq:parti_inhomo} 
as 
\begin{eqnarray}
Z\left\{\delta \hat{\rho}_p\right\} &=& e^{-F_h/k_B T}\int D[\tilde{c}(\mathbf{q})]
\exp \left [-S\left\{\tilde{c}(\mathbf{q})\right\}\right]\label{eq:weak_inhomo}
\end{eqnarray}
where the action $S$ is given by
\begin{eqnarray}
S\left\{\tilde{c}(\mathbf{q})\right\} &=& 
  \frac{V^2}{2n_p N}\sum_{\mathbf{q}\neq 0}\frac{\tilde{c}(\mathbf{q})\tilde{c}(-\mathbf{q})}{g_o(\mathbf{q})} + \frac{V}{2\rho_{p0}^2}\sum_{\mathbf{q}\neq 0}\left[\rho_{s0}^2\bar{J}^{-1}(\mathbf{q})-2\bar{\chi}_{ps}\right]\tilde{c}(\mathbf{q})\tilde{c}(-\mathbf{q}) \tilde{h}_p(\mathbf{q}) \tilde{h}_p(-\mathbf{q})\nonumber \\
&& + \frac{4\pi l_{Bo} z_p^2 \alpha_p^2 V}{2}\sum_{\mathbf{q}\neq 0} \frac{\tilde{c}(\mathbf{q})\tilde{c}(-\mathbf{q})\tilde{h}_p(\mathbf{q}) \tilde{h}_p(-\mathbf{q})}{m(\mathbf{q})}
- \frac{(4\pi l_{Bo})^2 z_p^4 w_{cr}^2}{4}\sum_{\mathbf{q}\neq 0}\tilde{c}(\mathbf{q})\tilde{c}(-\mathbf{q})\tilde{U}(\mathbf{q})\nonumber \\
&& + \frac{1}{2}\sum_{\mathbf{q}\neq 0}\ln \left[\frac{m(\mathbf{q})}{\mathbf{q}^2 \left[1 + \Delta \tilde{\epsilon}_h(\mathbf{q})\right] + \tilde{\kappa}^2(\mathbf{q})} \right] \label{eq:action_wip}
\end{eqnarray}
so that $m(\mathbf{q})$ is given by Eq. ~\ref{eq:mfactor} and Eq. 
~\ref{eq:inhomo_ft} becomes
\begin{eqnarray}
\tilde{g}(\mathbf{q}_1,\mathbf{q}_2,\mathbf{q}_3) &=& \frac{4\pi l_{Bo}}{3}\left[p_p^2\left\{\sum_{\gamma' = c,B^+}\frac{\beta_{\gamma'}}{e^{-\psi_{p,b}}}\right\}
\tilde{h}_p(\mathbf{q}_1)\tilde{h}_p(\mathbf{q}_2) \right . \nonumber \\
&& \left . - \frac{p_s^2\rho_{s0}}{\rho_{p0} \tilde{h}_s(\mathbf{q}_3)}\tilde{h}_s(\mathbf{q}_1)\tilde{h}_s(\mathbf{q}_2)\tilde{h}_p(\mathbf{q}_3)\right]\tilde{c}(\mathbf{q}_3)
\label{eq:inhomo_ft_final}
\end{eqnarray}
Also, $g_o(\mathbf{q}) = N g_D(q^2 Nb^2/6)$ so that $g_D(x) = 2\left[e^{-x}-1+x\right]/x^2$. 

\subsubsection{Dipolar polymers in an electrolyte solution}\label{app:H}
Consider a case when $\beta_{c} = 1, \beta_{B^+} = 0$ i.e., none of the dissociable groups on the chains dissociate and hence, none of the $B^+$ ions can absorb on the chains. In this case, the chains have electric dipoles on the backbones and the solution contains $B^+$ and $A^-$ ions in the solvent. This means, $\alpha_p = w_{cr} = \psi_{p,b} = 0$ (cf. Eqs. ~\ref{eq:para1}, ~\ref{eq:para2} and ~\ref{eq:para3}). Also, $n_{c}^f = (1-\beta_{c})n_p N =0$ and $n_{B^+}^f = n_{A^-} = n_{salt}$ for $z_{B^+} = -z_{A^-} = z$ so that (cf. Eqs. ~\ref{eq:mfactor},~\ref{eq:homo_ft}, ~\ref{eq:kappa_ft} and ~\ref{eq:inhomo_ft})
\begin{eqnarray}
m(\mathbf{q}_1) \equiv m_d(\mathbf{q}_1) &=& \mathbf{q}_1^2 (1 + \Delta\tilde{\epsilon}_{h,d}(\mathbf{q}_1)) + \tilde{\kappa}_d^2(\mathbf{q}_1) + \sum_{\mathbf{q}_3}\tilde{g}_d(\mathbf{q}_3,-\mathbf{q}_1,\mathbf{q}_3-\mathbf{q}_1)(\mathbf{q}_1\cdot \mathbf{q}_3)\label{eq:mfactor_dip}
\end{eqnarray}
and
\begin{eqnarray}
\Delta\tilde{\epsilon_h}(\mathbf{q}_1))\equiv \Delta\tilde{\epsilon}_{h,d}(\mathbf{q}_1)) &=& \frac{4\pi l_{Bo}p_s^2 n_s}{3V}\tilde{h}_s(\mathbf{q}_1)\tilde{h}_s(-\mathbf{q}_1) + \frac{4\pi l_{Bo}p_p^2 n_p N}{3 V}\tilde{h}_p(\mathbf{q}_1)\tilde{h}_p(-\mathbf{q}_1)\label{eq:homo_ft_dip}\\
\tilde{\kappa}^2(\mathbf{q}_1) \equiv \tilde{\kappa}_d^2(\mathbf{q}_1)&=& 4\pi l_{Bo} z^2 \frac{n_{salt}}{V} \sum_{\gamma = B^+,A^-}\tilde{h}_\gamma(\mathbf{q}_1)\tilde{h}_\gamma(-\mathbf{q}_1)\label{eq:kappa_ft_dip}
\end{eqnarray}
\begin{eqnarray}
\tilde{g}_d(\mathbf{q}_1,\mathbf{q}_2,\mathbf{q}_3) &=& \frac{4\pi l_{Bo}}{3}\left[p_p^2 \rho_{p0}
\tilde{h}_p(\mathbf{q}_1)\tilde{h}_p(\mathbf{q}_2)\tilde{h}_s(\mathbf{q}_3) - p_s^2\rho_{s0}\tilde{h}_s(\mathbf{q}_1)\tilde{h}_s(\mathbf{q}_2)\tilde{h}_p(\mathbf{q}_3)\right]\frac{\tilde{c}(\mathbf{q}_3)}{\rho_{p0}\tilde{h}_s(\mathbf{q}_3)}\nonumber \\
&& \label{eq:inhomo_ft_final_dip}
\end{eqnarray}
The free energy of the weakly inhomogeneous electrolyte solutions containing the dipolar chains ($F_d$) can be written using Eq. ~\ref{eq:weak_inhomo} as
\begin{eqnarray}
e^{-F_d/k_B T} &=& e^{-F_{h,d}/k_B T}\int D[\tilde{c}(\mathbf{q})]
\exp \left [-S_d\left\{\tilde{c}(\mathbf{q})\right\}\right]\label{eq:weak_inhomo_dip}
\end{eqnarray}
Here, $F_{h,d}/k_B T$ is the free energy of the homogeneous phase containing dipolar 
polymers in an electrolyte solution and can be readily obtained from Eq. ~\ref{eq:free_homo}. Action $S_d$ for the dipolar polymers can be obtained from Eq. ~\ref{eq:action_wip} and is given by
\begin{eqnarray}
S_d\left\{\tilde{c}(\mathbf{q})\right\} &=&
  \frac{V^2}{2n_p N}\sum_{\mathbf{q}\neq 0}\frac{\tilde{c}(\mathbf{q})\tilde{c}(-\mathbf{q})}{g_o(\mathbf{q})} + \frac{V}{2\rho_{p0}^2}\sum_{\mathbf{q}\neq 0}\left[\rho_{s0}^2\bar{J}^{-1}(\mathbf{q})-2\bar{\chi}_{ps}\right]\tilde{c}(\mathbf{q})\tilde{c}(-\mathbf{q}) \tilde{h}_p(\mathbf{q})\tilde{h}_p(-\mathbf{q})\nonumber \\
&& + \frac{1}{2}\sum_{\mathbf{q}\neq 0}\ln \left[\frac{m_d(\mathbf{q})}{\mathbf{q}^2 \left[1 + \Delta \tilde{\epsilon}_{h,d}(\mathbf{q})\right] + \tilde{\kappa}_d^2(\mathbf{q})} \right]
\end{eqnarray}
Expanding the logarithmic term in $S_d$ in powers of $\bar{c}$ and retaining up to quadratic terms, inverse of the structure factor can be readily identified. Here, 
we consider the case of spherically symmetric molecules of equal sizes so that 
$\tilde{h}_j(\mathbf{q}) = \tilde{h}(\mathbf{q}) = \exp\left[-q^2a^2/2\pi\right]$. Defining $\Delta_{ps} = 4\pi l_{Bo} \left[p_p^2 \rho_{p0} - p_s^2\rho_{s0}\right]/3$, we get
\begin{eqnarray}
S_d\left\{\tilde{c}(\mathbf{q})\right\} &=&
  \frac{V^2}{2n_p N}\sum_{\mathbf{q}\neq 0}\frac{\tilde{c}(\mathbf{q})\tilde{c}(-\mathbf{q})}{g_o(\mathbf{q})} + \frac{V}{2\rho_{p0}^2}\sum_{\mathbf{q}\neq 0}\left[\rho_{s0}^2\bar{J}^{-1}(\mathbf{q})-2\bar{\chi}_{ps}\right]\tilde{c}(\mathbf{q})\tilde{c}(-\mathbf{q}) \tilde{h}(\mathbf{q})\tilde{h}(-\mathbf{q})\nonumber \\
&& + \frac{\Delta_{ps}}{2\rho_{p0}}\sum_{\mathbf{q}\neq 0}\eta_1(\mathbf{q})\tilde{c}(\mathbf{q}) - \frac{\Delta_{ps}^2}{4\rho_{p0}^2}\sum_{\mathbf{q}\neq 0} \sum_{\mathbf{q}'\neq 0}\tilde{c}(\mathbf{q})\eta_2(\mathbf{q},\mathbf{q}') \tilde{c}(\mathbf{q}')
\end{eqnarray}
where
\begin{eqnarray}
\eta_1(\mathbf{q}) &=& \sum_{\mathbf{q}_1\neq 0}\frac{\left[\mathbf{q}_1^2 + \mathbf{q}\cdot \mathbf{q}_1\right]\tilde{h}(\mathbf{q}_1)\tilde{h}(\mathbf{q}+\mathbf{q}_1)}{\mathbf{q}_1^2 \left[1 + \Delta \tilde{\epsilon}_{h,d}(\mathbf{q}_1)\right] + \tilde{\kappa}_d^2(\mathbf{q}_1)}\\
\eta_2(\mathbf{q},\mathbf{q}') &=& \sum_{\mathbf{q}_1\neq 0}\frac{\left[\mathbf{q}_1^2 + \mathbf{q}\cdot \mathbf{q}_1\right]\left[\mathbf{q}_1^2 + \mathbf{q}'\cdot \mathbf{q}_1\right]\tilde{h}^2(\mathbf{q}_1)\tilde{h}(\mathbf{q}+\mathbf{q}_1)\tilde{h}(\mathbf{q}' + \mathbf{q}_1)}{\left\{\mathbf{q}_1^2 \left[1 + \Delta \tilde{\epsilon}_{h,d}(\mathbf{q}_1)\right] + \tilde{\kappa}_d^2(\mathbf{q}_1)\right\}^2}
\end{eqnarray}
As $\tilde{h}(\mathbf{q})$ depends on the magnitude of $\mathbf{q}$ ($=q$), we can integrate out the angular degrees of freedom in the 
continuum limit after writing $\sum_{\mathbf{q}} \equiv V\int d\mathbf{q}/(2\pi)^3$. Also, 
$\eta_2(\mathbf{q},-\mathbf{q}) \equiv \eta_2(q,-q)$ is required for 
calculation of the structure factor. The angular integrations lead to 
\begin{eqnarray}
\eta_1(q)&=& \frac{V}{2\pi^2}\exp\left[-q^2a^2/2\pi\right]\int_0^{\infty} dq_1 q_1^2 \frac{\sinh q_1 q a^2/\pi}{q_1 q a^2/\pi}\left[q_1^2 + \frac{\pi}{a^2}\left\{1 - \frac{q_1 q a^2}{\pi} \coth \frac{q_1 q a^2}{\pi} \right\}\right]\nonumber \\
&& \frac{\exp\left[-q_1^2a^2/\pi\right]}{q_1^2 \left[1 + \Delta \tilde{\epsilon}_{h,d}(q_1)\right] + \tilde{\kappa}_d^2(q_1)}\\
\eta_2(q,-q) &=& V\left[\tilde{\eta}_{2,0} - q^2 \tilde{\eta}_{2,1}\right]\exp\left[-q^2a^2/\pi\right] \\
\tilde{\eta}_{2,0} &=& \frac{1}{2\pi^2} \int_0^{\infty} dq_1 \frac{q_1^6 \exp\left[-2q_1^2a^2/\pi\right]}{\left[q_1^2 \left[1 + \Delta \tilde{\epsilon}_{h,d}(q_1)\right] + \tilde{\kappa}_d^2(q_1)\right]^2}\label{eq:L20}\\
\tilde{\eta}_{2,1} &=& \frac{1}{6\pi^2} \int_0^{\infty} dq_1 \frac{q_1^4 \exp\left[-2q_1^2a^2/\pi\right]}{\left[q_1^2 \left[1 + \Delta \tilde{\epsilon}_{h,d}(q_1)\right] + \tilde{\kappa}_d^2(q_1)\right]^2}\label{eq:L21}
\end{eqnarray}

In the limit of $a\rightarrow 0$, $\eta_1(q)$ becomes independent of $q$ and the sum 
$\sum_{\mathbf{q}\neq 0} \tilde{c}(\mathbf{q})$ vanishes due to the fact that $\int d\mathbf{r}\delta\hat{\phi}_p(\mathbf{r}) = 0$. So, $S_d$ can be written as
\begin{eqnarray}
S_d\left\{\tilde{c}(\mathbf{q})\right\} &=& \frac{V}{2}\int \frac{d\mathbf{q}}{(2\pi)^3}\tilde{c}(\mathbf{q})S_{dd}^{-1}(q)\tilde{c}(-\mathbf{q})\label{eq:action_dd}\\
S_{dd}^{-1}(q) &=& \frac{1}{<\tilde{c}(\mathbf{q})\tilde{c}(-\mathbf{q})>} \nonumber \\
&=&
  \frac{V^2}{n_p N}\frac{1}{g_o(q)} + \left\{\frac{V}{\rho_{p0}^2}\left[\rho_{s0}^2\bar{J}^{-1}(q)-2\bar{\chi}_{ps}\right]- \frac{\Delta_{ps}^2 V}{2\rho_{p0}^2}\left[\tilde{\eta}_{2,0} -q^2 \tilde{\eta}_{2,1}\right]\right \}\exp\left[-q^2a^2/\pi\right] \nonumber \\
&& \label{eq:invsdd}
\end{eqnarray}
where $\bar{J}^{-1}(q) = V\exp\left[q^2a^2/\pi\right]/n_s$. A similar correlation 
function for the melts can be obtained from Eq. ~\ref{eq:invsdd} by \textcolor{black}{replacing  
$\frac{1}{\rho_{p0}^2}\left[\rho_{s0}^2\bar{J}^{-1}(q)-2\bar{\chi}_{ps}\right]$ with $w_{pp}$, $w_{pp}$ being the excluded volume parameter, and using $p_s = 0$ for the melts. Furthermore, Eqs. ~\ref{eq:action_dd}-~\ref{eq:invsdd} lead to Eq. ~\ref{eq:heuristic_heff} via inverse Fourier transform.} Comparing Eq. ~\ref{eq:invsdd} with Eq. ~\ref{eq:heuristic_dip_st} for the melts, 
we can identify 
\begin{eqnarray}
\lambda_0 &=& - \frac{2}{\tilde{\eta}_{2,0}} \sim - a^3\\
\lambda_1 &=& \frac{2}{\tilde{\eta}_{2,1}} \sim a
\end{eqnarray}

\subsubsection{Charge regulating polyelectrolyte chains in an electrolyte solution}\label{app:I}
The free energy of the weakly inhomogeneous electrolyte solutions containing the charge regulating polyelectrolyte chains ($F_p$) can be
written using Eq. ~\ref{eq:weak_inhomo}
\begin{eqnarray}
e^{-F_p/k_B T} &=& e^{-F_h/k_B T}\int D[\tilde{c}(\mathbf{q})]
\exp \left [-S_p\left\{\tilde{c}(\mathbf{q})\right\}\right]\label{eq:weak_inhomo_pe}
\end{eqnarray}
where
\begin{eqnarray}
S_p\left\{\tilde{c}(\mathbf{q})\right\} &=& \frac{V}{2}\int \frac{d\mathbf{q}}{(2\pi)^3}\tilde{c}(\mathbf{q})S_{pp}^{-1}(q)\tilde{c}(-\mathbf{q})\\
S_{pp}^{-1}(q) &=& S_{dd}^{-1}(q) + \frac{4\pi l_{Bo} z_p^2 \alpha_p^2 V\tilde{h}(q)\tilde{h}(-q)}{q^2 \left[1 + \Delta \tilde{\epsilon}_h(q)\right] + \tilde{\kappa}^2(q)}
- \frac{(4\pi l_{Bo})^2 z_p^4 w_{cr}^2}{2}\tilde{U}(q)\label{eq:invstruct}
\end{eqnarray}
and we have used $\tilde{h}_j(\mathbf{q}) \equiv \tilde{h}(\mathbf{q}) \exp(-q^2 a^2/2\pi)$.
Here, $S_{dd}^{-1}(q)$ is given by Eq. ~\ref{eq:invsdd} with the substitutions of $\Delta \tilde{\epsilon}_{h,d}$ by $\Delta \tilde{\epsilon}_h$ and
$\tilde{\kappa}_d^2$ by $\tilde{\kappa}^2$ in Eqs. ~\ref{eq:L20} and ~\ref{eq:L21}. Also, $\Delta_{ps} = 4\pi l_{Bo} \left[p_p^2 \rho_{p0}\left\{\sum_{\gamma' = c,B^+}\frac{\beta_{\gamma'}}{e^{-\psi_{p,b}}}\right\} - p_s^2\rho_{s0}\right]/3$ in Eq. ~\ref{eq:invsdd}.
We can evaluate $\tilde{U}(\mathbf{q}) \equiv \tilde{U}(q)$ by expanding the denominator in 
Eq. ~\ref{eq:capM} in powers of $\tilde{\delta \phi}_p$. Also, for the structure factor, we 
need the term which is independent of $\tilde{\delta \phi}_p$. This, in turn, means that we need to evaluate (cf. Eq. ~\ref{eq:capM})
\begin{eqnarray}
\tilde{U}(\mathbf{q}) &=& \sum_{\mathbf{q}_1} \frac{\exp\left[-q_1^2 a^2/\pi - (\mathbf{q} - \mathbf{q}_1)^2 a^2/\pi\right]}{\left[q_1^2 \left[1 + \Delta \tilde{\epsilon}_h(q_1)\right] + \tilde{\kappa}^2(q_1)\right] \left[(q-q_1)^2 \left[1 + \Delta \tilde{\epsilon}_h(q-q_1)\right] + \tilde{\kappa}^2(q-q_1)\right]}\label{eq:capM_struct} \\
&=& \tilde{h}(q)\tilde{h}(-q)V \tilde{M}(\mathbf{q})
\end{eqnarray} 
In the continuum limit, for the case of $a=0$, writing the sum over $\mathbf{q}_1$ as an integral and evaluating the integral 
\begin{eqnarray}
\tilde{M}(\mathbf{q}) \equiv \tilde{M}(q)  &=& \frac{1}{16 q \left[1 + \Delta \tilde{\epsilon}_h(0)\right]^2} \left[1-\frac{2}{\pi}\arccos\left(\frac{q}{\sqrt{q^2 + \tilde{\kappa}^2(0))}}\right) \right . \nonumber \\
&& \left . + \frac{2}{\pi} \arctan\left(\frac{q^3}{\tilde{\kappa}(0)\left\{3q^2 + 4\tilde{\kappa}^2(0)\right\}} \right) \right] \label{eq:capM_salty}
\end{eqnarray} 
For salt-free (i.e., $n_{A^-} = n_{B^+} = 0$) polyelectrolyte melts, Eq. ~\ref{eq:invstruct} is identical to Eq. ~\ref{eq:heuristic_heff_charge}.

\subsection{Numerical evaluation of the structure factor for the melts of salt-free ionic polymers}
\begin{figure}[h]
  \includegraphics[width=0.65\linewidth]{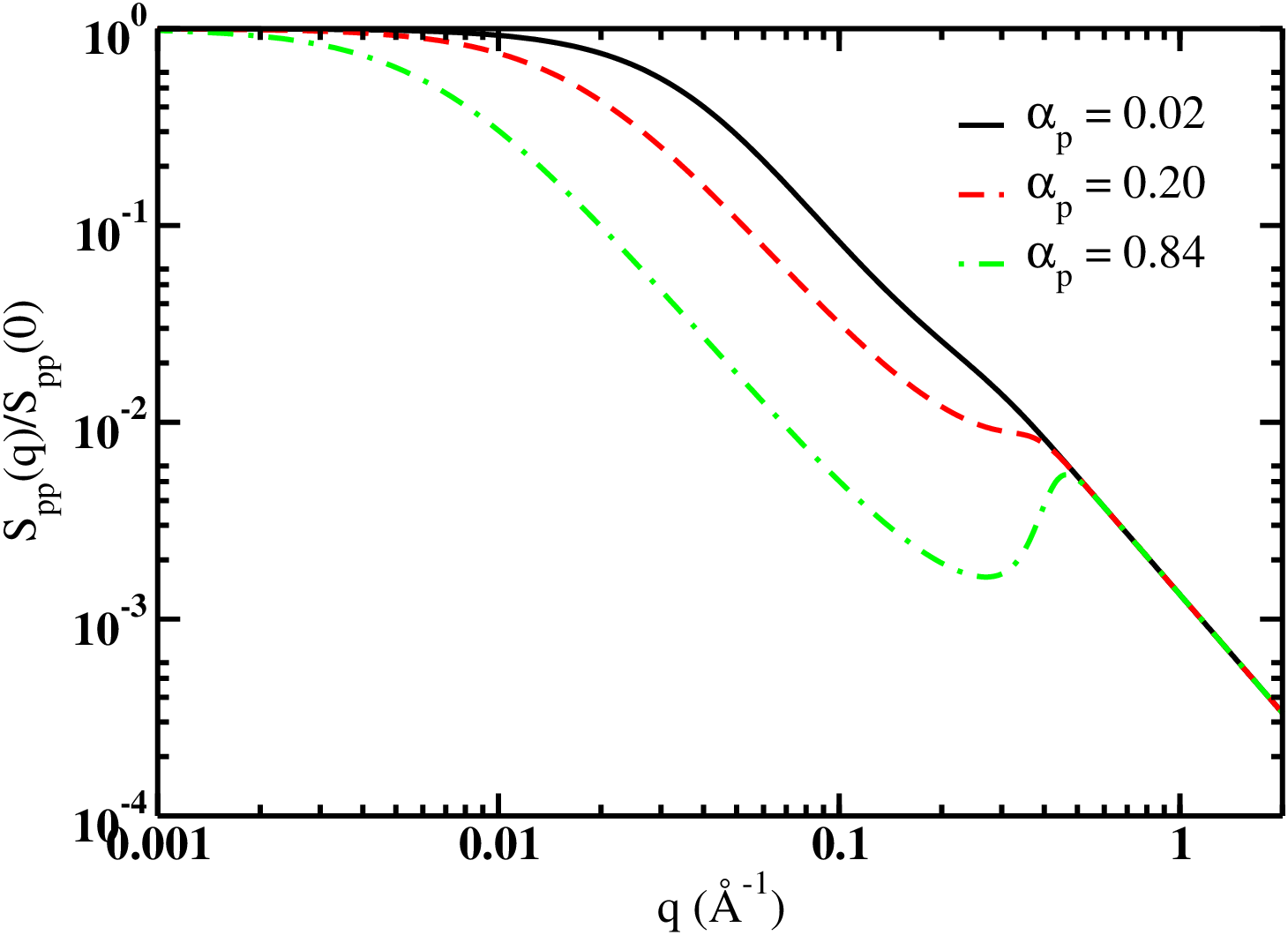}
  \includegraphics[width=0.65\linewidth]{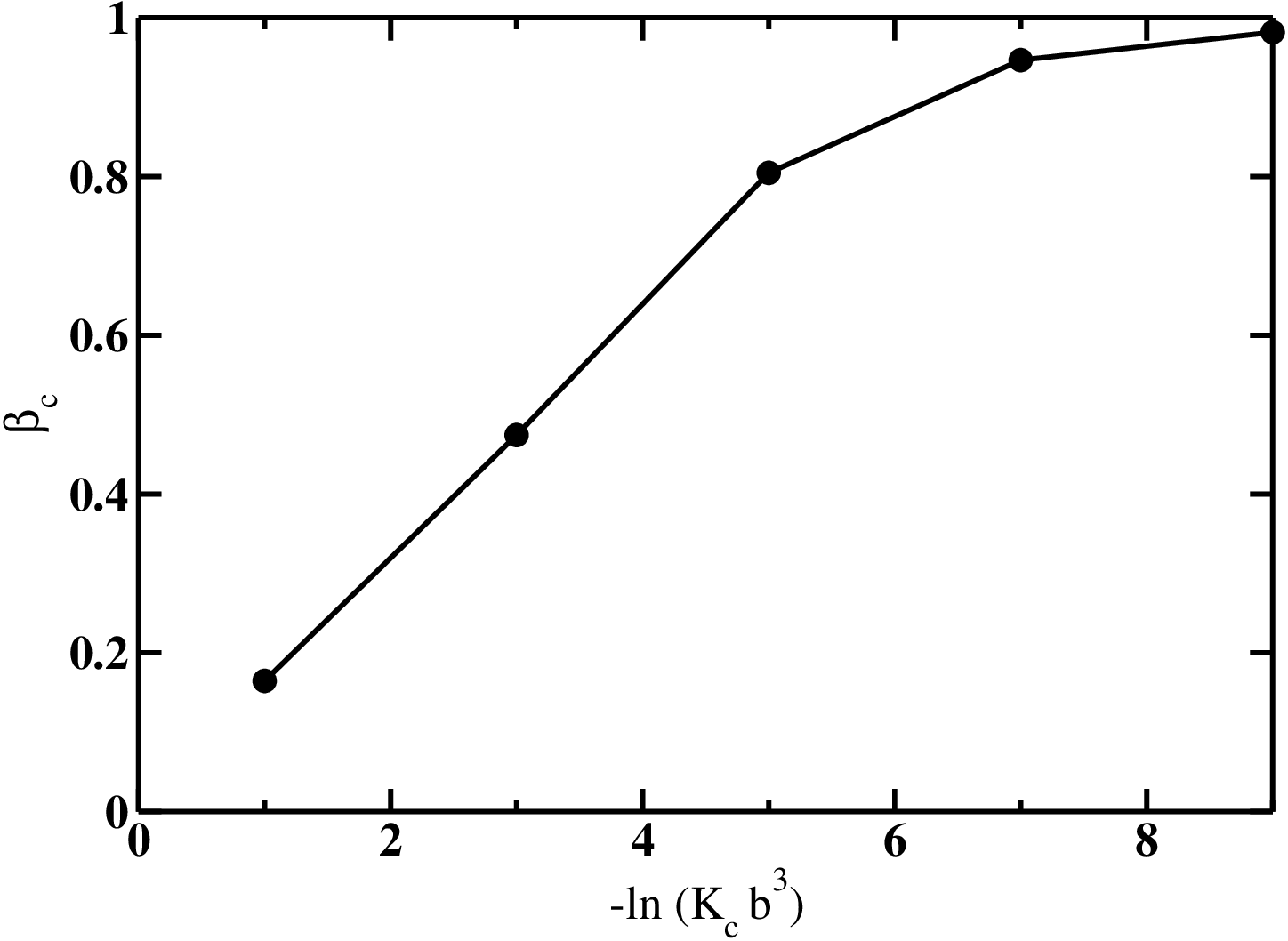}
\caption{(Top) Effects of the degree of ionization ($\alpha_p$) on the structure factor of melts containing charged polymers. (Bottom) Dependence of the degree of adsorption ($\beta_c = 1 -\alpha_p$) on the binding energy of the counterion
parameterized using the equilibrium constant, $K_c$. These curves were obtained by using
$T = 298 K, \psi_b = 0, a = 10 \mbox{\AA}, p_p = b = 3 \mbox{\AA}, \rho_{po} = n_p N/V =  10^{-3} \mbox{\AA}^{-3}, N = 1000$ and $V^2 S_{pp}(0)/n_p N = 10^{3}$.\label{fig:dipolar_effects}}
\vspace{0.2in}
\end{figure}

Eq. ~\ref{eq:invstruct} is general and can be applied to any charged or dipolar polymeric system. In Fig. ~\ref{fig:dipolar_effects}, we have used Eq. ~\ref{eq:invstruct} to highlight the
effects of counterion adsorption on the structure factor of melts i.e., in the absence of any solvent. In particular, it is shown that
changes in the degree of ionization, resulting from changes in the binding energy of the counterion-monomer pairs (see the bottom panel in Fig. ~\ref{fig:dipolar_effects}), can have significant effects on the
structure factor. An increase in the degree of ionization of monomers along the chains can lead to a peak in the structure factor at a finite wavevector (see the top panel in Fig. ~\ref{fig:dipolar_effects}) along with an upturn at lower wavevectors characterized by a minimum in the structure factor. This so-called polyelectrolyte peak appears as a result of interplay between charge-charge correlations along the chains and chain connectivity. The origin of the upturn has already been discussed in section ~\ref{sec:heuristic} and lies in the non-local effects of polarization. In the absence of the non-local effects of the polarization, it can be readily shown that either the structure factor decreases monotonically or a peak at a finite wavecector can appear in the structure factor\cite{borue_1988,joanny_1990,vilgis_rpa,liverpool_counterions}. However, the peak and the upturn at low wavevectors don't appear simultaneously in the absence of the dipolar interactions.

For computing the structure factor shown in Fig. ~\ref{fig:dipolar_effects}, we  have considered monovalent negatively charged monomers so that $z_p = -z_c = -1$. Also, the degree of adsorption $\beta_c$ was obtained by minimization of the free energy \textcolor{black}{with respect to $\beta_c$}, explicitly written as
\begin{eqnarray}
\frac{F }{n_p N k_B T} &=& \frac{F_a}{n_p N k_B T} + \frac{z_p^2 (1-\beta_c)e^{-2z_p\psi_b}}{2\left[z_c^2 e^{-2\psi_{p,b}} + z_p^2 \beta_c e^{-z_p\psi_b}\right]} \nonumber \\
&+& \frac{V}{4\pi^2n_p N}\int_0^{\infty}dq q^2 \ln \left[\frac{S_{pp}^{-1}(q)}{\hat{S}_{pp}^{-1}(q)} \left\{1 + \Delta \tilde{\epsilon}_h(q) + \frac{\tilde{\kappa}^2(q)}{q^2}\right\}\right] \label{eq:free_energy}
\end{eqnarray}
so that
\begin{eqnarray}
\frac{F_a}{n_p N k_B T} &=& \beta_c \ln \beta_c + 2(1-\beta_c)\ln \left[1-\beta_c\right] + \psi_{p,b} - (1-\beta_c)\ln \left[\frac{K_c e^{1-z_c\psi_b}V}{n_p N}\right]
\end{eqnarray}
, where $\psi_b$ is the electrostatic potential in the solution and
$\exp\left[-\psi_{p,b}\right] = \beta_{c} + \left(1-\beta_{c}\right) \exp\left[-z_p\psi_{b}\right]$.
$K_c$ is the equilibrium constant, formally defined by $K_c = e^{\mu_{pc}^o -\mu_{p}^o - \mu_{c}^o}$ so that
$\mu_{j}^o$ is the chemical potential of $j$ in an isolated state. In writing Eq. ~\ref{eq:free_energy}, the system containing the same number of chains but without any interactions among the components was taken as a reference.
This led to $\hat{S}_{pp}^{-1}(q)$ appearing in Eq. ~\ref{eq:free_energy}, which is the inverse structure factor for the solutions or melts containing the chains of the same lengths but in the absence of any interactions. Explicitly, $\hat{S}_{pp}^{-1}(q) = V^2/(n_p N g_o(q))$. Also, we have assumed that the characteristic size scale of a solvent molecule, a counterion and a monomer to be 
identical i.e., $a_c = a_s = a_p = a$. Furthermore, rather than varying $w_{pp}$  directly in Eq. ~\ref{eq:invsdd} (after replacing $\frac{1}{\rho_{p0}^2}\left[\rho_{s0}^2\bar{J}^{-1}(q)-2\bar{\chi}_{ps}\right]$ by $w_{pp}$ and using $p_s = 0$ in Eq. ~\ref{eq:invsdd}), we have computed the structure factor for a fixed value of the scattering intensity at $q=0$ to highlight non-monotonic nature of the structure facctor in weakly inhomogeneous phase. Numerical minimization of the free energy given in Eq. ~\ref{eq:free_energy} with respect to $\beta_c$ was done using Brent's method\cite{numerical_recipes}. Integrals in Eq. ~\ref{eq:free_energy} were evaluated using the Gauss-Legendre quadrature\cite{numerical_recipes} with $512$ points.   

\subsection{Comparisons with experiments}
\begin{figure}
  \includegraphics[width=1.0\linewidth]{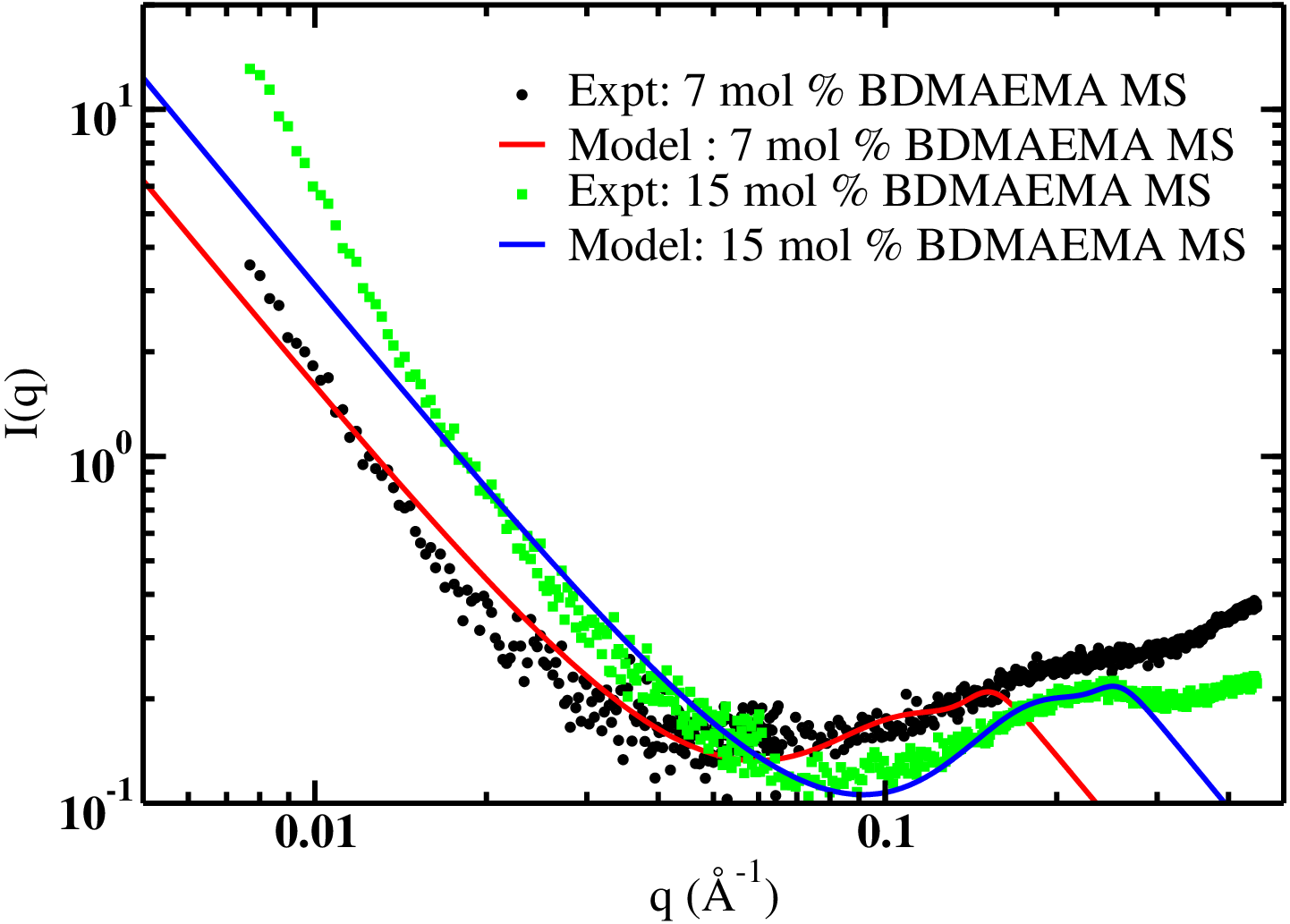}
\caption{Best fits of X-ray scattering data from Ref. \cite{long_paper} for random copolymers of n-butyl acrylate and 2-$[$butyl(dimethyl)-amino$]$ethyl methacrylate methanesufonate (BDMAEMA MS) containing different percentages of the charged moeities, BDMAEMA MS. The fits were obtained using Eq. ~\ref{eq:rayleigh_rewrite}. \label{fig:compare_expt_model}}
\vspace{0.2in}
\end{figure}

With the well-known limitations of the RPA for charged polymers\cite{Muthukumar2016,muthu_perspective}, it is prudent to expect Eq. ~\ref{eq:invstruct} to be applicable for concentrated solutions and melts of ionic polymers. In order to show the usefulness of Eq. ~\ref{eq:invstruct} for interpretting experimental data, we present Fig. ~\ref{fig:compare_expt_model} showing
the best fits of experimental data using Eq. ~\ref{eq:invstruct}. The experimental data represent X-ray scattering traces for random copolymers of n-butyl acrylate and charged 2-$[$butyl(dimethyl)-amino$]$ethyl methacrylate methanesufonate (BDMAEMA MS) monomers, which was taken from Ref. \cite{long_paper}.

For comparisons with the experiments, we rewrote Eq. ~\ref{eq:rayleigh} in a form
\begin{eqnarray}
I(q) &=& I(0)\frac{S_{pp}(q)}{S_{pp}(0)} = \frac{I(0)}{\tilde{h}(q)\tilde{h}(-q) + \frac{V^2}{n_p N}S_{pp}(0)\Delta S(q)}\label{eq:rayleigh_rewrite} 
\end{eqnarray}
and took $I(0)$ as a fitting parameter. Here, we have defined 
\begin{eqnarray}
\Delta S(q) &=& \frac{n_p N}{V^2}\left[\frac{1}{S_{pp}(q)} - \frac{\tilde{h}(q)\tilde{h}(-q)}{S_{pp}(0)}\right]
\end{eqnarray}
and used Eq. ~\ref{eq:weak_inhomo_pe}, which leads to $\left<\tilde{c}(\mathbf{q})\tilde{c}(-\mathbf{q})\right> = S_{pp}(q)$. Explicitly, 
\begin{eqnarray}
\frac{n_p N}{V^2 S_{pp}(q)} &=& \frac{n_p N}{V^2 S_{dd}(q)} + \left[\frac{4\pi l_{Bo} z_p^2 \alpha_p^2 n_p N/V}{q^2 \left[1 + \Delta \tilde{\epsilon}_h(q)\right] + \tilde{\kappa}^2(q)}
- \frac{(4\pi l_{Bo})^2 z_p^4 w_{cr}^2}{2}\frac{n_p N}{V}\tilde{M}(q)\right]\tilde{h}(q)\tilde{h}(-q) \nonumber \\
&&
\end{eqnarray}
which leads to 
\begin{eqnarray}
\frac{n_p N}{V^2 S_{pp}(0)} &=& \frac{n_p N}{V^2 S_{dd}(0)} + \frac{4\pi l_{Bo} z_p^2 \alpha_p^2 n_p N/V}{\tilde{\kappa}^2(0)}
- \frac{(4\pi l_{Bo})^2 z_p^4 w_{cr}^2}{2}\frac{n_p N}{V}\tilde{M}(0)
\end{eqnarray}
For dipolar polymer melts, substituting $\frac{1}{\rho_{p0}^2}\left[\rho_{s0}^2\bar{J}^{-1}(q)-2\bar{\chi}_{ps}\right] \equiv w_{pp}$ and $p_s = 0$, $\Delta S(q)$ becomes
\begin{eqnarray}
\Delta S(q) &=& \frac{1}{N}\left[\frac{1}{g_D(q^2 Nb^2/6)}-\exp\left[-q^2 a^2/\pi\right]\right] + \frac{8\pi^2 l_{Bo}^2p_p^4 n_p N}{9a V} q^2 \eta_{2,1}\exp\left[-q^2a^2/\pi\right]
\end{eqnarray} 
where $\eta_{2,1} = a \tilde{\eta}_{2,1}$ (cf. Eq. ~\ref{eq:L21}).

Fits for the X-ray scattering data were obtained by varying $I(0), V^{2}S_{pp}(0)/(n_p N^2), Nb^2/6, a$ and $\frac{8\pi^2 l_{Bo}^2p_p^4 n_p N}{9a V}\eta_{2,1}$. 
Self-consistent calculation of $\eta_{2,1}$ by numerically evaluating the integral in Eq. ~\ref{eq:L21} was avoided to expedite the fitting process. For the charged polymers, degree of adsorption ($\beta_c$), length of an ion-pair, $p_p$, and $n_p N/V$ were taken as additional parameters. 
The upper bound on $\beta_c$ was taken as the percentage of dissociable groups on each chain. The degree of polymerizations were estimated 
from the molecular weights and were kept fixed. In particular, $N=861$ and $N=1756$ were used for the sample with 7 mole $\%$ and 15 mole $\%$ BDMAEMA MS, respectively. Also, a Kuhn segment length ($b$) of 3 \AA  was assumed for both the samples and $T = 298$ K was used in computing the structure factor.
  
\begin{table}[h!]
\centering
\caption{Best fit parameters corresponding to the fits presented in Fig. 4 for the charged polymers.}
\vspace{0.3in}
\begin{tabular}{|c|c|c|c|c|c|c|c|}
\hline
{Sample} & $I(0)$ & $V^{2}S_{pp}(0)/(n_p N^2)$& %
    $\beta_c$ & $a$ (\AA) & $p_p$ (\AA) & $N\eta_{2,1}$ & $n_p N/V$ (\AA$^{-3}$)\\
\cline{1-8}
7 $\%$
BDMAEMA MS & 99627.00 & 27632.90 & 0.97 &  35.03 & 2.09 & 124.20 & 0.08\\
\cline{1-8}
15 $\%$ 
BDMAEMA MS & 9635.99 & 469.48 & 0.94 &  21.24  & 1.77 &354.93 &0.15\\
\cline{1-8}
\end{tabular}
\vspace{0.2in}
\label{table:1}
\end{table}

Fig. ~\ref{fig:compare_expt_model} shows that both the peak as well as the upturn in the scattering seen in the experiments can be fitted using Eq. ~\ref{eq:rayleigh_rewrite}. However, the scattering at even lower and higher wavevectors can not be described using Eq. ~\ref{eq:rayleigh_rewrite}. Failures of Eqs. ~\ref{eq:invstruct} and ~\ref{eq:rayleigh_rewrite} at even lower and higher wavevectors to fit the experimental data highlight need to improve the theory. Discrepancies at lower wave-vectors can be removed either by going beyond RPA\cite{muthu_double_screening,fredbook,riggleman_fts} or by accounting for possible aggregation. Consideration of aggregation along with the current model can indeed describe the scattering intensity at lower wavevectors\cite{Muthukumar2016}. \textcolor{black}{Recent simulation work\cite{chremos_jack} based on coarse-grained molecular dynamics simulations, which considered solvation of counterions and resulting aggregation (``void'' formation as per Ref. \cite{chremos_jack}), is another way of interpreting scattering at lower wave-vectors.} Discrepancies at higher wave-vectors highlight the lack of atomistic deails in the model developed in this work. \textcolor{black}{For example, the scattering intensity is predicted\cite{rayleigh_scattering, einstein_paper,debye_scattering} to decay like $q^{-4}$ in the limit of large wave-vector, while, in contrast, the intensity 
increases near the highest $q$ probed in Fig. ~\ref{fig:compare_expt_model}. The increase in intensity with $q$ in Fig. ~\ref{fig:compare_expt_model} hints at additional structure at shorter 
length scales. The model developed here can be used in a complementary manner with simulations, which capture atomistic details\cite{paddison_2016}, to fit scattering data with larger range of wave-vectors.} Systematic studies are needed in order to understand and get rid of the remaining discrepancies between the predicted scattering and experimental results. Nevertheless, the ability of Eq. ~\ref{eq:invstruct} to describe the peak as well as the upturn in the scattering at the same time \textit{purely on the basis of electrostatics} is unprecedented and the main development of this work.

\begin{figure}
  \includegraphics[width=1.0\linewidth]{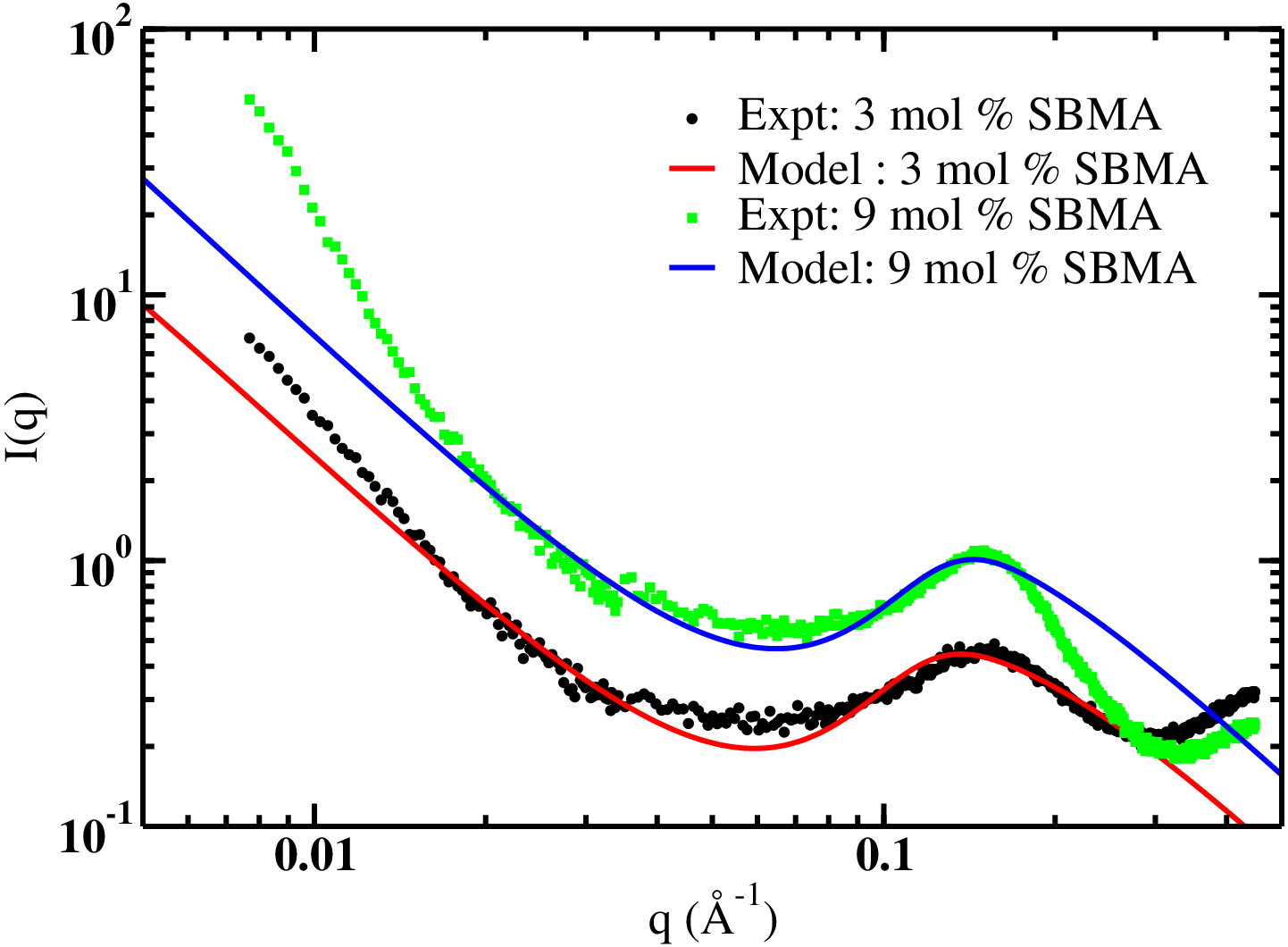}
\caption{Best fits for X-ray scattering data of the random copolymer melts containing n-butyl acrylate and zwitterionic 3-$[[$2-(methacryloyloxy)ethyl$]$(dimethyl)-ammonio$]$-1-propanesulfonate (SBMA) (data taken from Ref. \cite{long_paper}). The fits were obtained using Eq. ~\ref{eq:rayleigh_rewrite}  with $\beta_c = 1$. \label{fig:compare_expt_model_zwitt}}
\vspace{0.2in}
\end{figure}

\begin{table}[h!]
\centering
\caption{Best fit parameters corresponding to the fits presented in Fig. 5 for zwitterionic polymers.}
\vspace{0.3in}
\begin{tabular}{|c|c|c|c|c|c|}
\hline
{Sample} & $I(0)$ & $V^{2}S_{pp}(0)/(n_p N^2)$& %
    $\sqrt{Nb^2/6} (\AA)$ & $a$ (\AA) & $\frac{8\pi^2 l_{Bo}^2p_p^4 n_p N}{9a V}\eta_{2,1} (\AA^2)$\\
\cline{1-6}
3 mol $\%$ SBMA & 153.40 & 218.81 & 8.35 &  32.18 & 81617.40\\
\cline{1-6}
9 mol $\%$ SBMA & 3304.16 & 1702.06 & 9.76 &  29.39 & 74274.90\\
\cline{1-6}
\end{tabular}
\vspace{0.2in}
\label{table:2}
\end{table}

Similar fits of the scattering from dipolar polymer melts (i.e., when $\beta_c = 1$) are shown in Fig. ~\ref{fig:compare_expt_model_zwitt}, for the random copolymer melts of n-butyl acrylate and zwitterionic 3-$[[$2-(methacryloyloxy)ethyl$]$(dimethyl)-
ammonio$]$-1-propanesulfonate (SBMA) monomers, again taken from Ref. \cite{long_paper}. These fits demonstrate that the peak and the upturn in the scattering intensity can be 
described in dipolar polymer melts by introducing the non-local dielectric effects. Like the charged polymeric melts, the scattering at even lower and higher wavevectors can not be described using Eq. ~\ref{eq:rayleigh_rewrite} (see the case of 9 mole $\%$ SBMS in Fig. ~\ref{fig:compare_expt_model_zwitt}). These shortcomings of the current model again highlights needs to go beyond the RPA, accounting for possible aggregation and atomistic details while interpreting experimental results.

\section{Conclusions} \label{sec:conclusions}
We have considered the effects of electrostatic fluctuations for charge regulating polymers 
on the monomer-monomer density correlation function (structure factor). An analytical expression (Eq. ~\ref{eq:invstruct}) for the monomer-monomer structure factor was derived using the \textcolor{black}{random 
phase approximation}, valid for the concentrated solutions and melts. Comparisons with experimental data reveals that the peak as well as the upturn at lower wavevectors can be described using the analytical expression without considering any phase segregation or aggregation. Furthermore, consideration of electrostatic fluctuations has been shown to cause an oscillatory radial distribution function and induce attraction between similarly charged monomers. In addition, electrostatic interactions are shown to be screened in such a manner that the screening length has additional contributions from the fluctuating charges on the polymer backbones. In the future, we plan to extend the theory by going beyond the \textcolor{black}{random phase approximation} using variational methods\cite{muthu_double_screening} as well as field theoretic simulations\cite{fredbook,riggleman_fts}. Exploration of the implications of these developments on the dynamics of concentration fluctuations\cite{edwardsbook} in concentrated solutions and melts of ionic and dipolar polymers is a key direction for future work. \textcolor{black}{Use of free energy expressions derived in this work in understanding phase separation in polymer electrolytes and polymer-colloid mixtures is another interesting area of research.}

It should be emphasized that in the absence of any absorption, scattered intensity of light, X-rays or neutrons from any material contains information about the inhomogeneities brought about either due to 
the molecular structure of scatterers or interactions/correlations among the scatterers\cite{einstein_paper}. 
Calculation of the scattered intensity can be done in two ways. One way is to consider interactions between the incident 
wave and the scatterers, and this leads to a molecular description of the energy loss (in a particular direction) resulting from the 
scattering. This way leads to an expression for a characteristic attentuation length (inverse of so-called ``turbidity'') over which the energy gets lost due to the scattering, in terms of molecular parameters such as refractive index of the solvent and shape of the scatterers in the case of light scattering\cite{rayleigh_scattering, debye_scattering,zimm_molecular, fixman_molecular}. 
However, it becomes very difficult to consider interactions of the scatterers in this approach, which hinders calculations of the 
scattered intensity at higher concentrations. The other way is to consider density 
fluctuations of the scatterers about a uniform phase and then relate the scattered intensity to the density 
fluctuations\cite{einstein_paper,debye_scattering,zimm_polymer_scattering,correlation_hole}. As we have been focusing on scattering by the monomers in concentrated solutions and melts, we have 
considered the method of density fluctuations (of the monomers) and then constructed the scattered intensity. Since we don't have to specify nature of the interactions between the incident 
wave and the scatterers, the results presented in this work can be readily used to interpret scattering by light, 
X-rays or neutrons.

\textcolor{black}{\section*{Supplementary Material}
See supplementary material for the derivation of results presented in section III A. }

\section*{ACKNOWLEDGMENTS}
This research was conducted at the Center for Nanophase Materials Sciences, which is a US Department of Energy Office of Science User Facility. RK acknowledges support from the Laboratory Directed Research and Development program at ORNL and discussions with Prof. Philip (Fyl) Pincus about the structure factor of polyelectrolyte solutions.

\renewcommand{\theequation}{A-\arabic{equation}}
  % redefine the command that creates the equation no.
  \setcounter{equation}{0}  % reset counter
\renewcommand{\theequation}{A-\arabic{equation}}
  % redefine the command that creates the equation no.                                                              
  \setcounter{equation}{0}  % reset counter       
\section*{APPENDIX A : Field theoretic representation for the electrostatics} \label{app:A}
Electrostatic contributions to the partition function (i.e., Eq. ~\ref{eq:final_particle_elec}) can be written in a field theoretic 
form by using the Hubbard-Statonovich
transformation\cite{fredbook} leading to
\begin{eqnarray}
\exp\left[-H_e\right] &=& \frac{1}{\zeta_{\psi}} \int D\left[\psi\right]\exp \left[-i\int d\mathbf{r}
\left\{\hat{\rho}_e(\mathbf{r}) - \nabla_{\mathbf{r}}.\hat{P}_{ave}(\mathbf{r})\right\} \psi(\mathbf{r}) +
\frac{1}{8\pi l_{Bo}}\int d\mathbf{r}\psi(\mathbf{r})\nabla_\mathbf{r}^2 \psi(\mathbf{r}))\right], \nonumber \\
&& \label{eq:hubba}
\end{eqnarray}
where
\begin{eqnarray}
\zeta_{\psi} &=& \int D\left[\psi\right]\exp \left[\frac{1}{8\pi
l_{Bo}}\int d\mathbf{r}\psi(\mathbf{r})\nabla_\mathbf{r}^2 \psi(\mathbf{r}))\right].
\end{eqnarray}
Using the transformation and Eq. ~\ref{eq:prob_dist}, we can integrate over
the orientations of the dipoles analytically and evaluate the average over $\theta_\alpha$.
Defining
integrals over the orientational degrees of freedom as $I$, we can write
\begin{eqnarray}
       I\left\{\psi,\mathbf{R}_\alpha,\theta_\alpha,\mathbf{r}_{k}\right\} & = & \int \prod_{\alpha=1}^{n_p}\prod_{t_\alpha=0}^{N} d\mathbf{u}_{\alpha}(t_\alpha) \int \prod_{k=1}^{n_s} d\mathbf{u}_{k}\exp \left[i\int d\mathbf{r}\left\{\nabla_{\mathbf{r}}.\hat{\textbf{P}}_{ave}(\mathbf{r})\right\} \psi(\mathbf{r})\right] \\
&=& I_p\left\{\psi,\mathbf{R}_\alpha,\theta_\alpha\right\}I_s\left\{\psi,\mathbf{r}_{k}\right\} \label{eq:orient_int}
\end{eqnarray}
where
\begin{eqnarray}
I_p\left\{\psi,\mathbf{R}_\alpha,\theta_\alpha\right\} = (4\pi)^{n_p N}\prod_{\alpha=1}^{n_p}\prod_{t_\alpha=0}^{N} \left[\frac{\sin\left\{p_p (1-\theta_{\alpha}(t_{\alpha}))\left|\int d\mathbf{r}\psi(\mathbf{r})\nabla_{\mathbf{r}}
\hat{h}_{p}(\mathbf{r}-\mathbf{R}_\alpha(t_{\alpha}))\right|\right\}}{p_p (1-\theta_{\alpha}(t_{\alpha}))\left|\int d\mathbf{r}\psi(\mathbf{r})\nabla_{\mathbf{r}}\hat{h}_{p}(\mathbf{r}-\mathbf{R}_\alpha(t_{\alpha}))\right|}\right]&&\nonumber \\
= (4\pi)^{n_p N}\exp \left[\int d\mathbf{r}' \hat{\phi}_p(\mathbf{r}')\ln \left[\frac{\sin\left\{p_p (1-\theta_{\alpha}(t_{\alpha}))\left|\int d\mathbf{r}\psi(\mathbf{r})\nabla_{\mathbf{r}}\hat{h}_
{p}(\mathbf{r}-\mathbf{r}')\right|\right\}}{p_p (1-\theta_{\alpha}(t_{\alpha}))\left|\int d\mathbf{r}\psi(\mathbf{r})\nabla_{\mathbf{r}}\hat{h}_
{p}(\mathbf{r}-\mathbf{r}')\right|}\right]\right] && \label{eq:orient_poly}
\end{eqnarray}
and we have defined
\begin{eqnarray}
\hat{\phi}_p(\mathbf{r}) &=& \sum_{\alpha=1}^{n_p}\int_{0}^N dt_{\alpha} \delta (\mathbf{r}-\mathbf{R}_\alpha(t_{\alpha})),
\end{eqnarray}
which is microscopic number density of the center of mass of the segments. 
Similarly,
\begin{eqnarray}
I_s(\psi,\mathbf{r}_{k}) &=& (4\pi)^{n_s}\exp \left[\int d\mathbf{r}' \hat{\phi}_s(\mathbf{r}')\ln \left[\frac{\sin\left\{p_s \left|\int d\mathbf{r}\psi(\mathbf{r})\nabla_{\mathbf{r}}\hat{h}_{s}(\mathbf{r}-\mathbf{r}')\right|\right\}}
{p_s \left|\int d\mathbf{r}\psi(\mathbf{r})\nabla_{\mathbf{r}}\hat{h}_
{s}(\mathbf{r}-\mathbf{r}')\right|}\right]\right]\label{eq:orient_solv}
\end{eqnarray}
and we have defined
\begin{eqnarray}
\hat{\phi}_s(\mathbf{r}) &=& \sum_{k=1}^{n_s}\delta (\mathbf{r}-\mathbf{r}_{k})
\end{eqnarray}
Using Eqs. ~\ref{eq:hubba}, ~\ref{eq:orient_int}, ~\ref{eq:orient_poly} and ~\ref{eq:orient_solv}, the
partition function given by Eq. ~\ref{eq:parti_sing} becomes
\begin{eqnarray}
       Z & = & \frac {\Lambda^{-3n'}}{n_s!n_p!n_{A^-}!\prod_{\gamma'}n_{\gamma'}^f}\int \prod_{\alpha=1}^{n_p}D[\mathbf{R}_\alpha]
\int \prod_{k=1}^{n_s}d\mathbf{r}_{k} \int \prod_{\gamma}\prod_{j=1}^{n_\gamma} d\mathbf{r}_{j}\exp \left[-H_0\left\{\mathbf{R}_\alpha\right\}
- H_w\left\{\mathbf{R}_\alpha,\mathbf{r}_{k}\right \} \right ]\nonumber \\
&& \int \frac{D\left[\psi\right]}{\zeta_\psi} \exp\left[ \frac{1}{8\pi l_{Bo}}\int d\mathbf{r}\psi(\mathbf{r})\nabla_\mathbf{r}^2 \psi(\mathbf{r})
 - i\int d\mathbf{r} \sum_{\gamma}z_\gamma \hat{\rho}_\gamma(\mathbf{r})\psi(\mathbf{r})\right]
I_s\left\{\psi,\mathbf{r}_{k}\right\} J_p(\psi,R_{\alpha})\nonumber \\
&& \prod_{\mathbf{r}}\mathbf{\delta}\left[\sum_{j=p,s,\gamma}\frac{\hat{\rho}_{j}(\mathbf{r})}{\rho_{j0}} -  1\right]\label{eq:hami_physical1}
\end{eqnarray}
where
\begin{eqnarray}
       J_p(\psi,R_{\alpha}) & = & \sum_{\left\{\theta_{\alpha}\right\}}\left< \exp\left[-E\left\{\theta_{\alpha}\right\} -i\int d\mathbf{r} z_p \hat{\rho}_{pe}(\mathbf{r})\psi(\mathbf{r})\right] I_p(\psi,R_{\alpha},\theta_\alpha) \right> \label{eq:jp_factor}
\end{eqnarray}
Eq. ~\ref{eq:jp_factor} can be readily evaluated using Eqs. ~\ref{eq:charge_sum}, ~\ref{eq:binding_energy}, ~\ref{eq:prob_dist}, ~\ref{eq:entropic_factor} and
~\ref{eq:orient_poly}, which gives
\begin{eqnarray}
       J_p(\psi,R_{\alpha}) & = & \Upsilon e^{-E}(4\pi)^{n_pN} \prod_{\alpha=1}^{n_p}\prod_{t_{\alpha}=0}^N\left[(1 - \beta_{c} - \beta_{B^+}) \exp\left[-i z_p \int d \mathbf{r} 
\hat{h}_p(\mathbf{r}-\mathbf{R}_{\alpha})\psi(\mathbf{r})\right]
\right . \nonumber \\
&& \left . + \left(\sum_{\gamma = c,B^+} \beta_{\gamma}\right) \left[\frac{\sin\left(p_p \left|\int d\mathbf{r}\psi(\mathbf{r})
\nabla_{\mathbf{r}}\hat{h}_{p}(\mathbf{r}-\mathbf{R}_{\alpha})\right|\right)}
{p_p \left|\int d\mathbf{r}\psi(\mathbf{r})\nabla_{\mathbf{r}}\hat{h}_
{p}(\mathbf{r}-\mathbf{R}_{\alpha})\right|} \right]\right] \\
&=& \Upsilon e^{-E} (4\pi)^{n_pN} \exp\left[-i \sum_{\alpha=1}^{n_p}\int_0^N dt_{\alpha}\int d\mathbf{r} \hat{h}_p(\mathbf{r}-\mathbf{R}_{\alpha})\psi_p(\mathbf{r})\right] \nonumber \\
&=& \Upsilon e^{-E} (4\pi)^{n_pN} \exp\left[-i \int d\mathbf{r} \hat{\rho}_p(\mathbf{r})\psi_p(\mathbf{r})\right] \label{eq:jpfactor_approx}
\end{eqnarray}
Here, we have defined a quantity $\psi_p$ and $E \equiv E\left\{\theta_{\alpha}\right\}$ is given by Eq. ~\ref{eq:binding_energy}
with the relation $n_{\gamma'}^a = \beta_{\gamma'} n_p N$ for $\gamma' = c,B^+$. Using Eq. ~\ref{eq:jpfactor_approx}, Eq. ~\ref{eq:hami_physical1} can be written as
\begin{eqnarray}
       Z & = & e^{-F_a/k_B T}\int \prod_{\alpha=1}^{n_p} D[\mathbf{R}_\alpha]
\int \prod_{\gamma} \prod_{j=1}^{n_{\gamma}}
d\mathbf{r}_{j} \prod_{k=1}^{n_{s}} d\mathbf{r}_{k} \exp \left [-H_0\left\{\mathbf{R}_\alpha\right\}
- H_w\left\{\mathbf{R}_\alpha,\mathbf{r}_{k}\right \} \right ]\nonumber \\
&&  \int \frac{D\left[\psi\right]}{\zeta_\psi} \exp\left[\frac{1}{8\pi l_{Bo}}\int d\mathbf{r}\psi(\mathbf{r})\nabla_\mathbf{r}^2 \psi(\mathbf{r}))
 - i\int d\mathbf{r} \sum_{\gamma}z_\gamma \hat{\rho}_\gamma(\mathbf{r})\psi(\mathbf{r}) - i \int d\mathbf{r} \hat{\rho}_p(\mathbf{r})\psi_p(\mathbf{r}) \right . \nonumber \\
&& \left . + \int d\mathbf{r}' \hat{\phi}_s(\mathbf{r}')\ln \left[\frac{\sin\left\{p_s \left|\int d\mathbf{r}\psi(\mathbf{r})\nabla_{\mathbf{r}}\hat{h}_{s}(\mathbf{r}-\mathbf{r}')\right|\right\}}
{p_s \left|\int d\mathbf{r}\psi(\mathbf{r})\nabla_{\mathbf{r}}\hat{h}_
{s}(\mathbf{r}-\mathbf{r}')\right|}\right] \right]
\prod_{\mathbf{r}}\mathbf{\delta}\left[\sum_{j=p,s,\gamma}\frac{\hat{\rho}_{j}(\mathbf{r})}{\rho_{j0}} -  1\right]
\label{eq:parti_elec_added} 
\end{eqnarray}

where
\begin{eqnarray}
       \frac{F_a}{k_B T} & = & n_{B^+}^a\ln \mbox{K}_{B^+} - (n_pN-n_{c}^a)\ln \mbox{K}_{c}
- \ln \left[\frac{n_pN!}{n_{c}^a! n_{B^+}^a!(n_pN- n_{c}^a - n_{B^+}^a)!}\right] \nonumber \\
&& + \ln \left[n_{c}^f! n_{B^+}^f! n_{A^-}!n_s!n_p!\right] - (n_pN + n_s) \ln 4\pi + 3n'\ln \Lambda
\end{eqnarray}

We rewrite Eq. ~\ref{eq:parti_elec_added} in a form given by Eq. ~\ref{eq:parti_elec_integrated} after writing the local incompressibility condition as a funcional integral using the identity
\begin{eqnarray}
\prod_{\mathbf{r}}\mathbf{\delta}\left[\sum_{j=p,s,\gamma}\frac{\hat{\rho}_{j}(\mathbf{r})}{\rho_{j0}} -  1\right] &=& 
\int D\left[\eta\right] \exp\left[-i \int d\mathbf{r} \eta(\mathbf{r})\left\{\sum_{j=p,s,\gamma}\frac{\hat{\rho}_{j}(\mathbf{r})}{\rho_{j0}} -  1 \right\}\right]\label{eq:pressure_field}
\end{eqnarray}
and defining partition functions for the individual ions after integrating over 
their positions. 

\bibliography{structure_factor_revised_3}
\end{document}